\documentclass[twocolumn]{revtex4-1}

\usepackage{bm}
\usepackage{comment}
\usepackage{amsmath}
\usepackage{amsfonts}
\usepackage{amssymb}
\usepackage{mathrsfs}

\usepackage{ulem}

\date{\today}

\usepackage[dvipdfmx]{graphicx}

\usepackage{color}

\newcommand{\dg}{\dagger}
\newcommand{\la}{\langle}
\newcommand{\ra}{\rangle}
\newcommand{\al}{\alpha}
\newcommand{\sg}{\sigma}
\newcommand{\gm}{\gamma}
\newcommand{\ep}{\varepsilon}

\begin{document}
\title{
Vortex bound state of Kondo lattice coupled to compensated metal
}
\author{Shoma Iimura$^{1}$} 
\author{Motoaki Hirayama$^2$}
\author{Shintaro Hoshino$^{1}$}

\affiliation{
$^1$Department of Physics, Saitama University, Shimo-Okubo, Saitama 338-8570, Japan\\
$^2$RIKEN Center for Emergent Matter Science (CEMS), Wako, Saitama 351-0198, Japan
}

\begin{abstract}
We theoretically study physical properties of the low-energy quasiparticle excitations at the vortex core in the full-gap superconducting state of the Kondo lattice coupled to compensated metals. 
Based on the mean-field description of the superconducting state, we numerically solve the Bogoliubov-de Gennes (BdG) equations for the tight-binding Hamiltonian.
The isolated vortex is characterized by a length scale 
independent of the magnitude of the interaction 
and the energy level of the core bound state is the same order as the bulk gap.
These properties are in strong contrast to the conventional $s$-wave superconductor.
To gain further insights, we also consider the effective Hamiltonian in the continuous limit and construct the theoretical framework of the quasiclassical Green's function of conduction electrons. 
With the use of the Kramer-Pesch approximation, we analytically derive the spectral function describing the quasiparticle excitations which is consistent with the numerics. 
It has been revealed that the properties of the vortex bound state are closely connected to the characteristic odd frequency dependence of both the normal and anomalous self-energies which is proportional to the inverse of frequency.
\end{abstract}

\maketitle

\section{Introduction}
Superconductors are classified into two types by the magnetic responses to the applied field \cite{Abrikosov57}. 
Type-I superconductor excludes the magnetic flux from the bulk and turns into the normal state at the critical field. 
On the other hand, the magnetic flux can penetrate into the type-II superconductor, where the superconducting order parameter becomes spatially non-uniform and the quantum vortices are formed. 
The physics of the superconducting vortex has been studied intensively \cite{Parks69}. 
For instance, 
the $s$-wave superconductor with the applied magnetic field forms the vortex and has the low-energy bound state known as Caroli-de Gennes-Matricon (CdGM) mode \cite{Caroli64}. 
The topologically non-trivial vortex bound state in the iron-based compounds has been studied for their application to the quantum computation \cite{Hosur11,Wang18,Machida19}. 
The quasiparticle excitation spectrum in real space can be observed by the recent advanced experiments such as scanning tunneling microscope measurements \cite{Hess89,Chen18}. 

Recently, motivated by the experiments that identify a full-gap nature of the superconducting states of CeCu$_2$Si$_2$ and UBe$_{13}$ \cite{Kittaka14,Shimizu15},
we have proposed a new mechanism of the full-gap superconductivity relevant to compensated metals interacting with the localized spin/pseudospin moments \cite{Iimura19}. 
This mechanism is associated with a frustration 
originating from the multichannel Kondo effects \cite{Emery92,Cox98}:
the over-screened localized moment mediates the quantum-mechanical superposition between the electron Fermi surface and the hole Fermi surface to form the Bogoliubov quasiparticle \cite{Iimura19}. 
The resultant U(1) symmetry breaking is characterized not by a conventional Cooper pair amplitude among the conduction electrons, but by a composite pair amplitude \cite{Emery92,Balatsky93,Coleman93,Coleman95,Coleman99,Flint08,Hoshino14}, which describes a three-body bound state involving itinerant electron, hole, and localized spin/pseudospin moment. 
We have studied the Meissner response to the uniform field and revealed that the magnetic penetration depth is longer than the usual BCS superconductor \cite{Iimura19}. 
In addition, we have also found that the uniform magnetic field induces the second-order transition \cite{Iimura19-2}, while the conventional BCS superconductor shows only the first-order transition at the Pauli limit. 
We thus expect that the physical properties of the low-energy quasiparticles within the vortex core are also different since the spin/pseudospin, which is described as an effective fermionic degrees of freedom, mediates the formation of the Bogoliubov quasiparticle.

In this paper, we study the low-energy properties of the isolated vortex in the Kondo lattice with compensated metallic conduction bands (CMCB-KL). 
For this purpose, we utilize the mean-field approximation \cite{Zhang00,Hoshino14-2,Iimura19} to describe the superconducting state. 
We consider the Kondo lattice with non-Kramers pseudospins, which has been suspected as an origin of some heavy-electron superconductors with non-Fermi liquid behavior \cite{Cox98}. 
We first discuss the tight-binding model numerically, and show characteristic properties of the vortex in the CMCB-KL; the length scale of the vortex bound state in the CMCB-KL is independent of the magnitude of the order parameter, 
in contrast to the BCS superconductors, where the length scale is proportional to the inverse of the superconducting gap function $\Delta$. 
The energy level spacing of the core states is the order of a bulk superconducting gap, and this point is also different from the BCS case where its level spacing is the order of $\Delta^2/E_{\rm F}$ with $E_{\rm F}$ ($\gg \Delta$) being the Fermi energy.
To further elucidate the low-energy properties of the CdGM mode, we derive the Eilenberger equation, which is the quasiclassical version of the Gor'kov-Dyson's equation. 
In the derivation process, we find that the superconducting electrons feel the self-energy inversely proportional to frequency, i.e., the odd-frequency superconductivity is realized in the CMCB-KL. 
With the use of the Kramer-Pesch approximation \cite{Kramer74,Pesch74}, which is used to analyze the vortex bound state in the BCS superconductor, 
we derive the energy dispersion of the CdGM mode in the CMCB-KL and reveal that the unique physical properties are associated with the characteristic frequency dependence of the self-energy. 

The rest of this paper is organized as follows.  
In Sec.\ref{sec:tight_binding_model}, we present numerical results for the tight-binding model, which are obtained by solving the Bogoliubov-de Gennes (BdG) equation for a finite-sized system self-consistently. 
In Sec.\ref{sec:effective_theory}, we introduce the effective Hamiltonian in the continuum limit of the tight-binding model 
to investigate the physical origin of the characteristics of the vortex. 
The summary of our work is given in Sec.\ref{sec:summary}.
In the following, we take $\hbar$, $k_{\rm B}$ and the lattice constant $a$ as unity.

\section{Tight-binding model}
\label{sec:tight_binding_model}
\subsection{Mean-Field Theory}
\label{sec:model}

We introduce the tight-binding model of the CMCB-KL. 
Focusing on the non-Kramers $\Gamma_3$ doublet as the ground state of the localized $f$-electron with $f^2$ configuration in cubic symmetry \cite{Cox98}, the Kondo lattice model is given by
\begin{align}
&{\mathcal H} ={\mathcal H}_0 + {\mathcal H}_{\rm int}, \nonumber \\
&{\mathcal H}_0 =  \sum_{\langle i,j \rangle \al \sg} t_{ij} \left( c_{i\alpha \sg}^\dagger c_{j\al\sg}^{ } + \mathrm{H.c.} \right) -\mu  \sum_{i\al\sg}  \sg^z_{\al \al} n_{i\al\sg}^{ }, \\
&{\mathcal H}_{\rm int} =  \frac{1}{2} J \sum_{i \al \al^\prime \sg} {\bm T}_i \cdot c_{i\al \sg}^\dagger {\bm \sg}_{\al \al^\prime} c_{i\al^\prime \sg}^{ },
\end{align}
where $c_{i \al \sg}$ ($c_{i \al \sg}^\dagger$) is an annihilation (creation) operator of the conduction electron. 
The corresponding particle number operator is given by $n_{i\al\sg} = c_{i\al\sg}^\dagger c_{i\al\sg}^{ }$. 
The index $i$ denotes the lattice site located at ${\bm R}_i$.
$\sg=\uparrow,\downarrow$ describes the Kramers indices associated with the time-reversal symmetry. 
$\alpha=1,2$ stands for the band index of the compensated metal. 
We have introduced the $2\times 2$ Pauli matrix ${\bm \sg}=(\sg^x,\sg^y,\sg^z)$.
${\bm T}_i$ is the local pseudospin moment describing the non-Kramers doublet and couples with the conduction electrons through the band index $\alpha$. 
$J$ is the coupling constant of the two-channel Kondo interaction.
We note that the symmetry of the Kondo-coupling  between $\sg=\uparrow$ and $\downarrow$ is preserved by the time-reversal symmetry \cite{Iimura19}. 
Therefore, the frustration on the Kondo-screening between spin-up electron and spin-down electron remains 
and can cause superconductivity.

\begin{figure}[t]
 \centering
    \includegraphics[width=88mm]{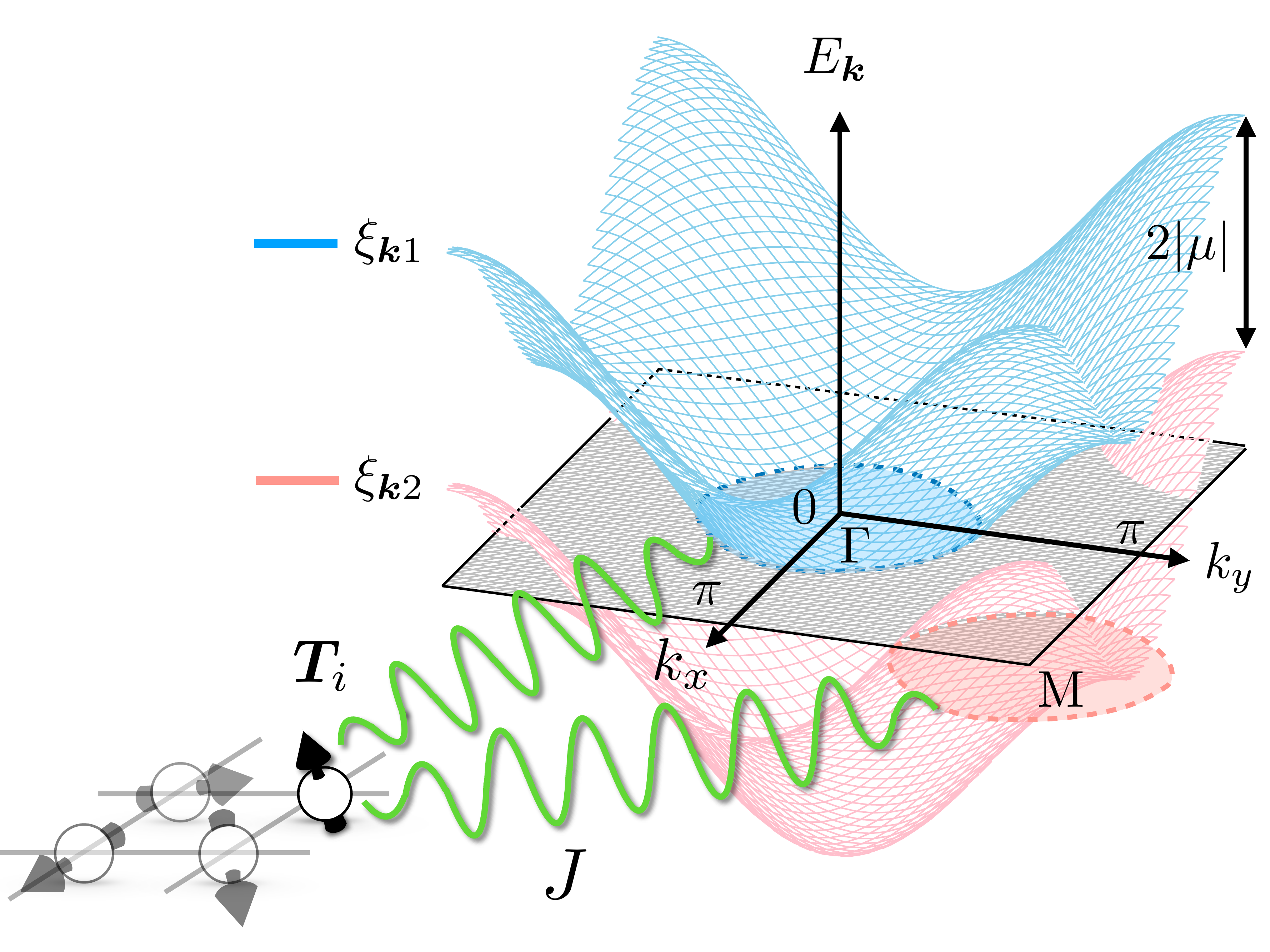}
  \caption{(Color online) Schematic picture of the tight-binding model. 
Blue and pink curved surfaces respectively represent the dispersion relation of $\xi_{{\bm k}1}$ and that of $\xi_{{\bm k}2}$. 
Grey plane represents the Fermi energy and the black border describes the Briilouin Zone (BZ) in the square lattice.
Blue and pink dotted circles on the grey plane respectively describe the electron Fermi surface and the hole Fermi surface. 
  }
  \label{fig:schematic_tb}
\end{figure}

Assuming that the vortex penetrates into the thin-film of the superconductor, we now consider the square lattice for simplicity. 
The spatial coordinate is written as ${\bm R}_i = (X_i,Y_i)$, where $X_i = i_x - (N_x-1)/2~(i_x=0,\cdots,N_x-1)$ and $Y_i = i_y - (N_y-1)/2~(i_y=0,\cdots,N_y-1)$. 
The number of the lattice sites is $N=N_x N_y$.
$t_{ij}$ is the hopping amplitude on the nearest neighbor bond in the square lattice. 
The onsite potential $\mu$ ($<0$) resolves the degeneracy between $\alpha=1,2$ to form the compensated metallic conduction bands. 
The single-particle energy dispersion of conduction electrons is then given by 
\begin{align}
	\label{eq:band_dispersion}
	\xi_{{\bm k}\al} = - 2 t \left( \mathrm{cos} k_x + \mathrm{cos} k_y \right) - \mu \sg^z_{\al \al},
\end{align}
In this setup, $\xi_{{\bm k}1}$ has an electron Fermi surface around $\Gamma$ point (${\bm k}=(0,0)$), while $\xi_{{\bm k}2}$ has a hole Fermi surface around the ${\rm M}$ point (${\bm k}=(\pi ,\pi)$) whose size is same as that of the electron Fermi surface. (See also Fig.\ref{fig:schematic_tb}.)
The Fermi energy $E_{\rm F}$ measured from the bottom of $\xi_{{\bm k}1}$ is 
expressed as $E_{\rm F}=4t + \mu$. 
The equivalence of the Fermi volume of the electron band and that of the hole band is always guaranteed in the case of integer fillings. 

We next introduce the mean-fields describing the superconducting state. 
To this end, we first rewrite the localized moment ${\bm T}_i$ in terms of the pseudofermion degrees of freedom $\{ f_{i1} , f_{i2} \}$, which is introduced as follows,
\begin{align}
	\label{eq:pseudo_fermion}
	&{\bm T}_i = \frac{1}{2} \sum_{\al \al^\prime} f_{i\al}^\dagger {\bm \sg}_{\al \al^\prime} f_{\al^\prime}.
\end{align}
with $\sum_\al f_{i\al}^\dagger f_{i\al} = 1$, which is the constraint on the localized pseudofermion at each site.
Assuming that this constraint is satisfied in average as $\sum_\al \la f_{i\al}^\dagger f_{i\al}\ra = 1$,  
then the interaction term can be decoupled in the mean-field approximation as follows,
\begin{align}
	\label{eq:hybridization_lattice}
	&{\mathcal H}_{\rm int} \simeq \sum_{i\al } \left( V^*_{i\al \uparrow} c^\dagger_{i\al \uparrow} + W^*_{i{\bar \al}\downarrow} \epsilon_{\al {\bar \al}} c_{i{\bar \al}\downarrow}^{ } \right) f_{i\al}^{ } + {\mathrm H.c.},
\end{align}
where ${\bar \al}$ is the complementary component of $\alpha$ such as ${\bar 1}=2$. 
${\hat \epsilon} = i {\hat \sigma}^y$ is the anti-symmetric unit tensor. 
The mean-field amplitudes $( V_{i\al\uparrow} , W_{i{\bar \al}\downarrow} )$ are defined as follows,
\begin{align}
	\label{eq:v}
	&V_{i\alpha \uparrow} = \frac{J}{4} \left( \langle f_{i\alpha} c_{i\al \uparrow}^\dagger \rangle + 2 \langle f_{i {\bar \al}}^{ } c_{i {\bar \al}\uparrow}^\dagger \rangle \right),\\
	\label{eq:w}
	&W_{i{\bar \alpha} \downarrow} = \frac{J}{4}\epsilon_{\al {\bar \al}} \left( \langle f_{i\alpha} c_{i{\bar \al} \downarrow}^{ } \rangle - 2 \langle f_{i {\bar \al}}^{ } c_{i \al \downarrow}^{~} \rangle \right).
\end{align}
We here assume the $s$-wave symmetry of the order parameters since the original Kondo coupling is local. 
To satisfy the constraint on the pseudofermion number, 
in general we need to add the potential term $\varepsilon_f \sum_{i\al} (f_{i\al}^\dagger f_{i\al} - 1 )$ with the Lagrange multiplier $\ep_f$ to the Hamiltonian. 
However, we can take $\varepsilon_f=0$ in the non-Kramers doublet systems due to the particle-hole symmetry of the superconducting state \cite{Iimura19}. 
Then, the BdG Hamiltonian in the CMCB-KL is given by
\begin{align}
\label{eq:BdG_Hamiltonian_lattice}
{\mathcal H}_{\rm BdG} & = \sum_{\al} {\vec{\Psi}_{\al}}^\dagger {\hat {\mathcal H}_{\al}} {\vec \Psi}_{\al}^{ } + \mathrm{const.},\\
{\hat {\mathcal H}}_\al &= \left( 
\begin{array}{ccc}
{\hat \xi}_{\al} & 0 & {\hat V}^\dagger_{\al \uparrow} \\
0 & -{\hat \xi}_{\bar \al}^T & {\hat W}^\dagger_{{\bar \al}\downarrow} \\
{\hat V}_{\al \uparrow} & {\hat W}_{{\bar \al}\downarrow} &0 
\end{array} 
\right),
\end{align}
where ${\vec \Psi}_{\al} = (\begin{array}{ccc} {\vec c}_{\al \uparrow} & {{\vec c}_{{\bar \al}\downarrow}}^{~\dagger} & {\vec f}_{\al} \end{array} )^T$ is the Nambu basis with ${\vec c}_{\al \sg} = (c_{1\al\sg},c_{2\al\sg},\cdots,c_{N\al \sg})^T$ and ${\vec f}_\al =(f_{1\al},f_{2\al},\cdots,f_{N\al})^T$. 
The matrix elements of each block matrix are then given by $({\hat \xi}_{\al})_{ij} = t_{ij} - \mu \sg^z_{\al \al} \delta_{ij}$, $({\hat V}_{\al \uparrow})_{ij} = V_{i\al\uparrow} \delta_{ij}$, and $({\hat W}_{{\bar \al}\downarrow})_{ij} = W_{i{\bar \al}\downarrow}\epsilon_{\al {\bar \al}} \delta_{ij}$. 

The conduction electrons are hybridized with the pseudofermions by the mean-fields, which effectively describe the formation of the heavy-fermion band \cite{Zhang00,Hoshino14-2,Iimura19}. 
This Fermi-liquid picture of the Kondo lattice is justified in the low-temperature limit, where the coherence among the local Kondo-clouds is activated well \cite{Capponi01}. 
Although the introduction of the mean-fields is asymmetric with respect to the spin indices, this fact is associated with the fact that the definition of the pseudofermion in Eq.\eqref{eq:pseudo_fermion} is not unique \cite{Hoshino14-2}. 
The symmetry between up and down spin indices is preserved if we evaluate the physical quantities in terms of the original physical degrees of freedom, 
i.e., the field operator of the conduction electrons and the pseudospin moment ${\bm T}_i$. 
Indeed, the superconducting state in the CMCB-KL is described by the formation of the composite pair amplitude \cite{Iimura19}
\begin{align}
\label{eq:composite}
\Phi_{i\al \al^\prime \sg \sg^\prime}
&\equiv \langle {\bm T}_i \cdot c_{i \al \sg} {\bm \sg}_{\al \al^\prime} c_{i
 \al^\prime \sg^\prime} \rangle,
\\
&\propto
V_{\al\uparrow}^* W_{{\bar \al}\downarrow} \epsilon_{\sg \sg^\prime} \epsilon_{\al \al^\prime}
\end{align}
which has symmetric form between spin-up and down. 
Therefore, the U(1) symmetry breaking is characterized by the coexistence of the mean-field amplitudes in Eqs.\eqref{eq:v} and \eqref{eq:w}.

\subsection{Numerical Results}
\label{sec:numerical_results}
\subsubsection{Method of numerical solution for isolated vortex}
We here solve Eqs.~\eqref{eq:v} and \eqref{eq:w} by iterative method to obtain the self-consistent solutions of the mean-field amplitudes. 
In order to consider a simple compensated metallic situation, where the electron Fermi surface and the hole Fermi surface are separated, we take $\mu/t = -3$ ($E_{\rm F}=t$).
We focus on the physical properties of the isolated vortex by assuming that the magnetic field is weak so that the distance between the penetrating vortices is large enough.

We choose the open boundary condition for the matrix elements of the hopping term in the BdG Hamiltonian. 
In addition, we consider the type-II limit with long enough London penetration depth, because the superconducting state in the CMCB-KL shows the magnetic penetration depth much larger than the usual value in the BCS theory \cite{Iimura19}. 
We then ignore the vector potential ${\bm A}$ to describe the vortex as the topological defect of the velocity potential of the superconducting electrons. 
In order to describe the single vortex, we use the non-uniform initial value for the mean-fields, which are given by
\begin{align}
&V_{i\al\uparrow} = |V_{i\al\uparrow}| \mathrm{e}^{i \theta_{i\al \uparrow}},\\
&W_{i{\bar \al}\downarrow} = |W_{i{\bar \al}\downarrow}| \mathrm{e}^{i \theta_{i{\bar \al} \downarrow}},
\end{align}
where $\theta_{i \al \sg} =\nu_{\al \sg} \mathrm{tan}^{-1} (Y_i/X_i)$ with the integer $\nu_{\al \sg} = 0 , \pm 1, \pm 2,\cdots$. 
$\mathrm{tan}^{-1} (Y_i/X_i)\equiv \varphi({\bm R}_i)$ is the azimuth angle in the two-dimensional polar coordinate systems. 
We note that the vorticity of the superconducting electrons is characterized by the difference ${\delta \nu}_\al =\nu_{\al\uparrow} - \nu_{{\bar \al}\downarrow}$ 
because the composite pair amplitude $\Phi_{i \al \al^\prime,\sg \sg^\prime}$ is decoupled into 
\begin{align}
	\Phi_{i \al \al^\prime,\sg \sg^\prime} \propto  |V_{i \al \uparrow}| | W_{i {\bar \al}\downarrow}|\mathrm{exp}[-i \delta \nu_\al \varphi({\bm R}_i)]
\end{align}
within the mean-field theory. 
Therefore, the physical properties of the superconducting state with $(\nu_{\al \uparrow},\nu_{{\bar \al}\downarrow})=(1,0)$ is equivalent to those with $(\nu_{\al \uparrow},\nu_{{\bar \al}\downarrow})=(0,-1)$. 
We therefore take $\nu_{\al\uparrow}=+1$ and $\nu_{{\bar \al}\downarrow}=0$ to describe the vortex with the single flux quanta in the following. 
As a criterion for convergence of the iteration, we use the threshold $10^{-4}$ for 
the relative error of the mean-fields at all the sites.
In addition to the vortex state, 
we will also consider the non-topological defect by introducing the impurity at the center of system for comparison.

\begin{figure*}[t]
 \begin{center}
    \includegraphics[width=93mm]{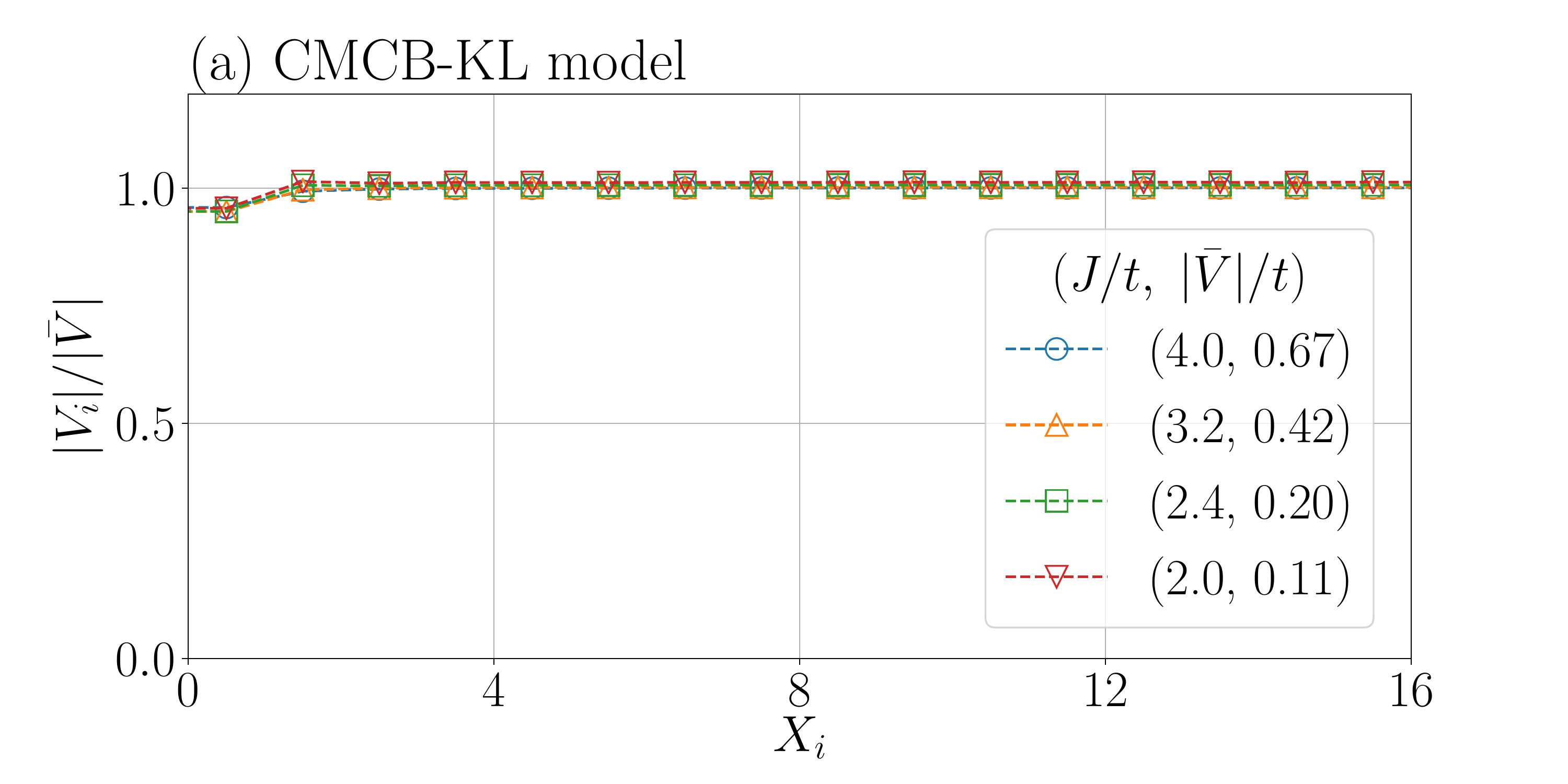}
    \hspace{-10mm}
    \includegraphics[width=93mm]{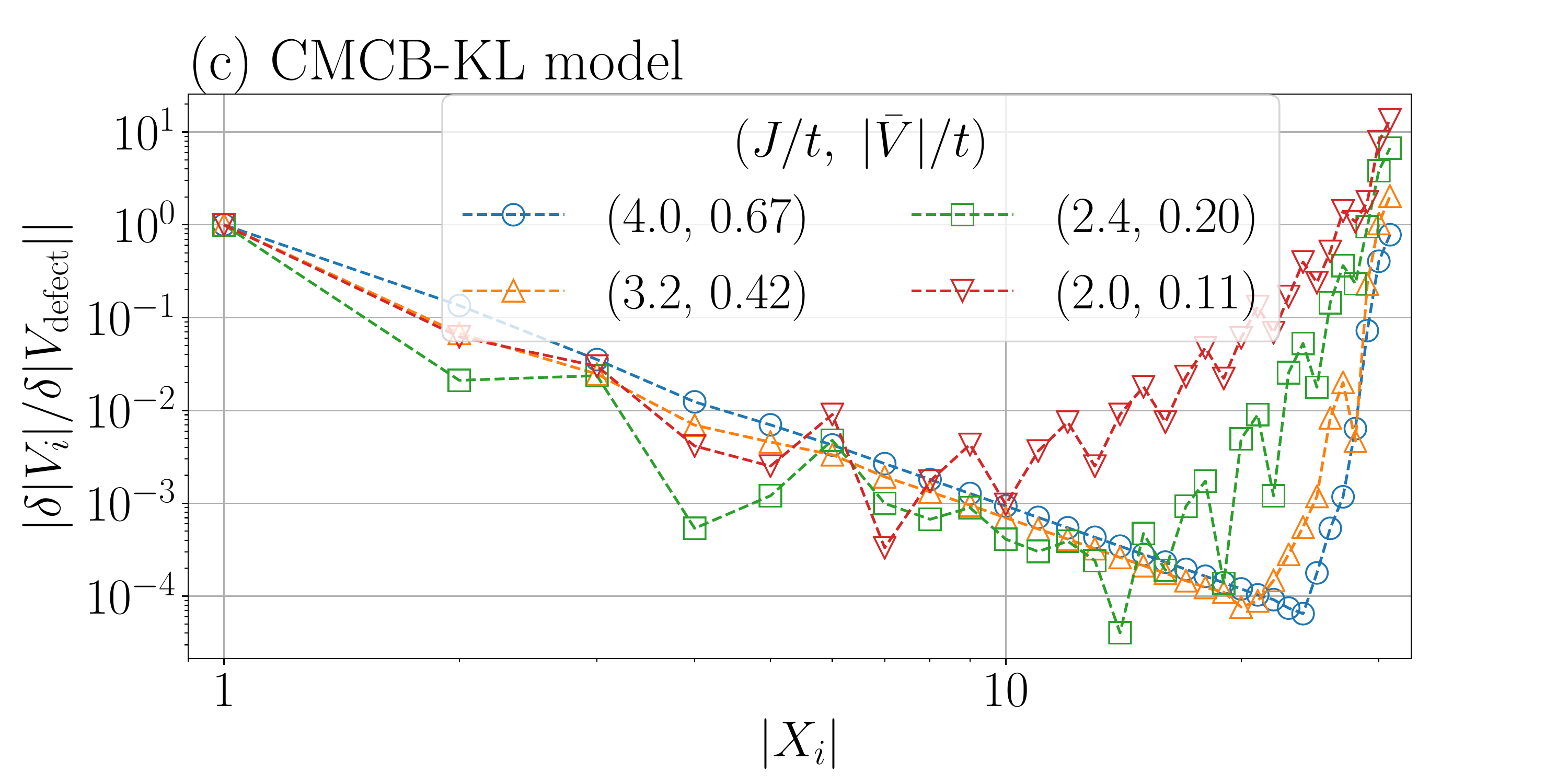}
\\
    \vspace{-2mm}
    \includegraphics[width=93mm]{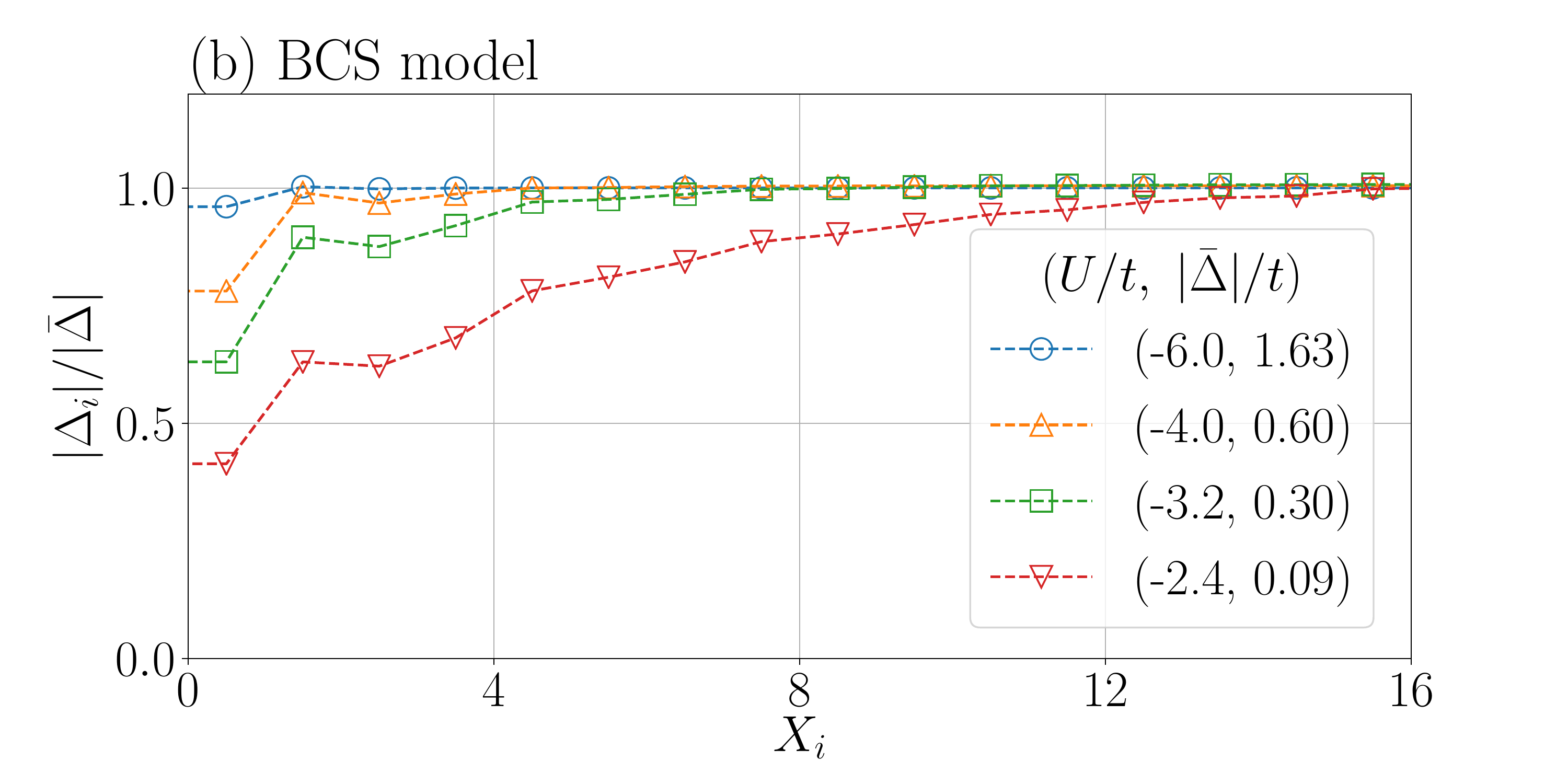}
    \hspace{-10mm}
    \includegraphics[width=93mm]{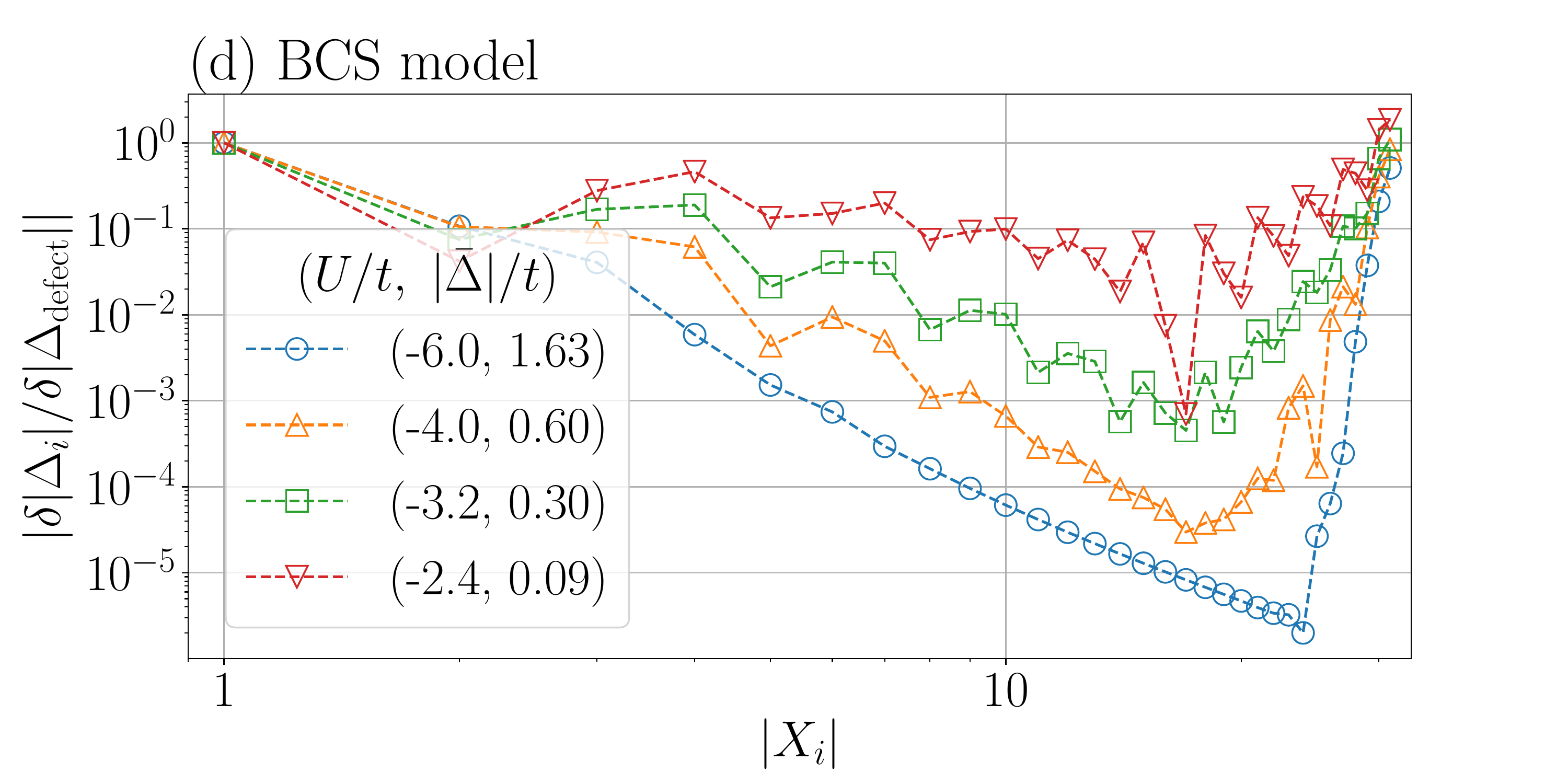}
  \caption{(Color online) Self-consistent solutions of the order parameters in the presence of the vortex for $N_x=N_y=64$. 
  In panels (a) and (b), the amplitudes of the order parameters are respectively normalized by the real-space average as
(a): $|{\bar V}|= \sum_{i} |V_{i}|/N$ and (b): $|{\bar \Delta}|=\sum_i |\Delta_i|/N$. 
Panels on the right-hand side show
(c) the derivative of the order parameters in the CMCB-KL 
$\delta |V (X_i + 1/2) | = |V(X_i + 1)| - |V(X_i)|$ and (d) that in the BCS model 
$\delta |\Delta (X_i +1/2 ) | = |\Delta(X_{i}+1)| - |\Delta(X_i)|$. 
  $X_i=0$ and $X_i=32$ respectively represent the coordinate of the center of the vortex and that of the boundary, which is parallel to the $y$-axis. 
  The markers respectively represent the parameters shown in the each panel. The Fermi energy is $E_{\rm F}/t=1$ and the temperature is $T=0$.
  }
  \label{fig:mean_fields_vortex}
 \end{center}
\end{figure*}

\subsubsection{Isolated vortex state as topological defect}

\begin{figure*}[t]
 \centering
    \includegraphics[width=93mm]{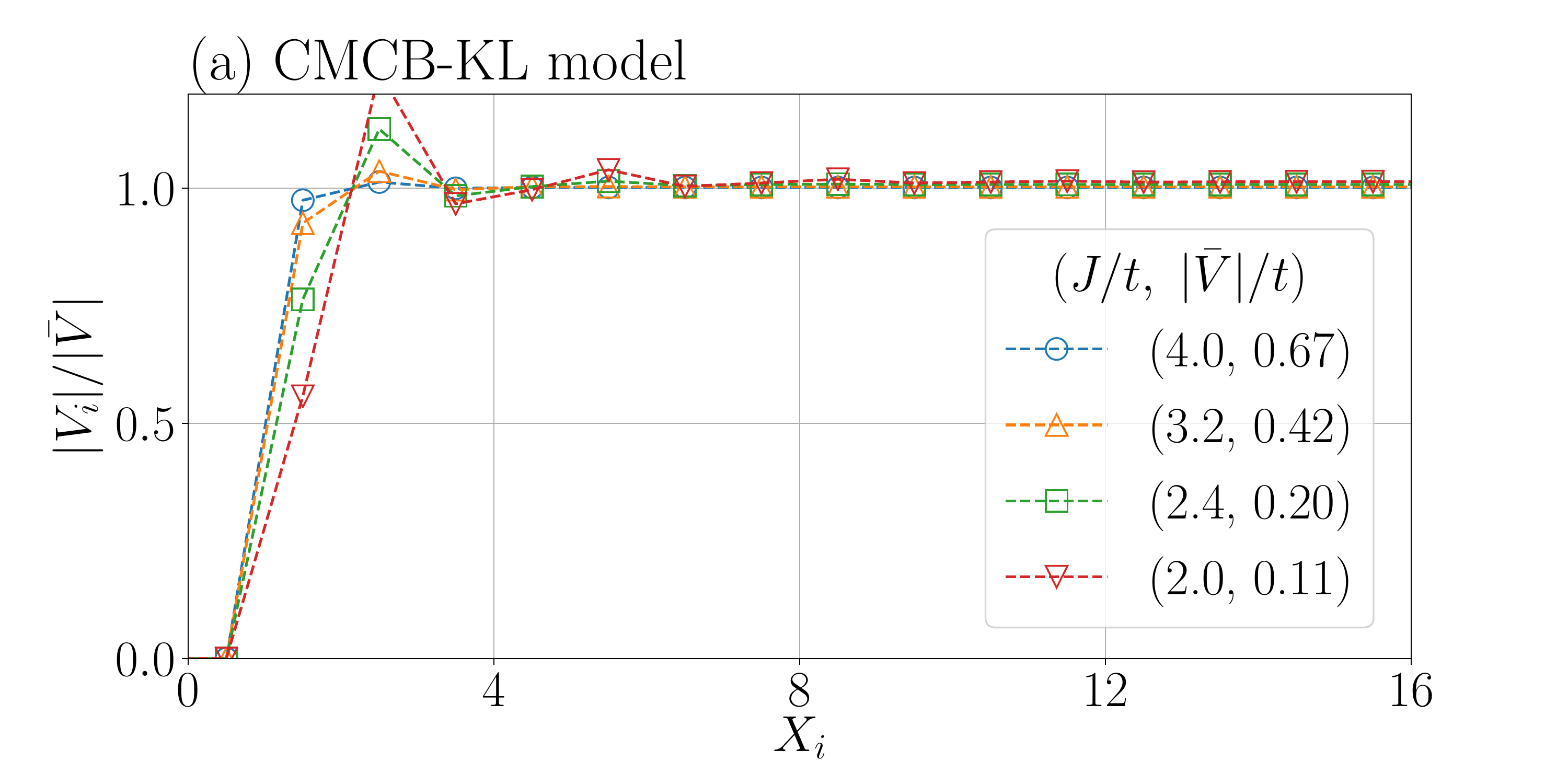}
    \hspace{-10mm}
    \includegraphics[width=93mm]{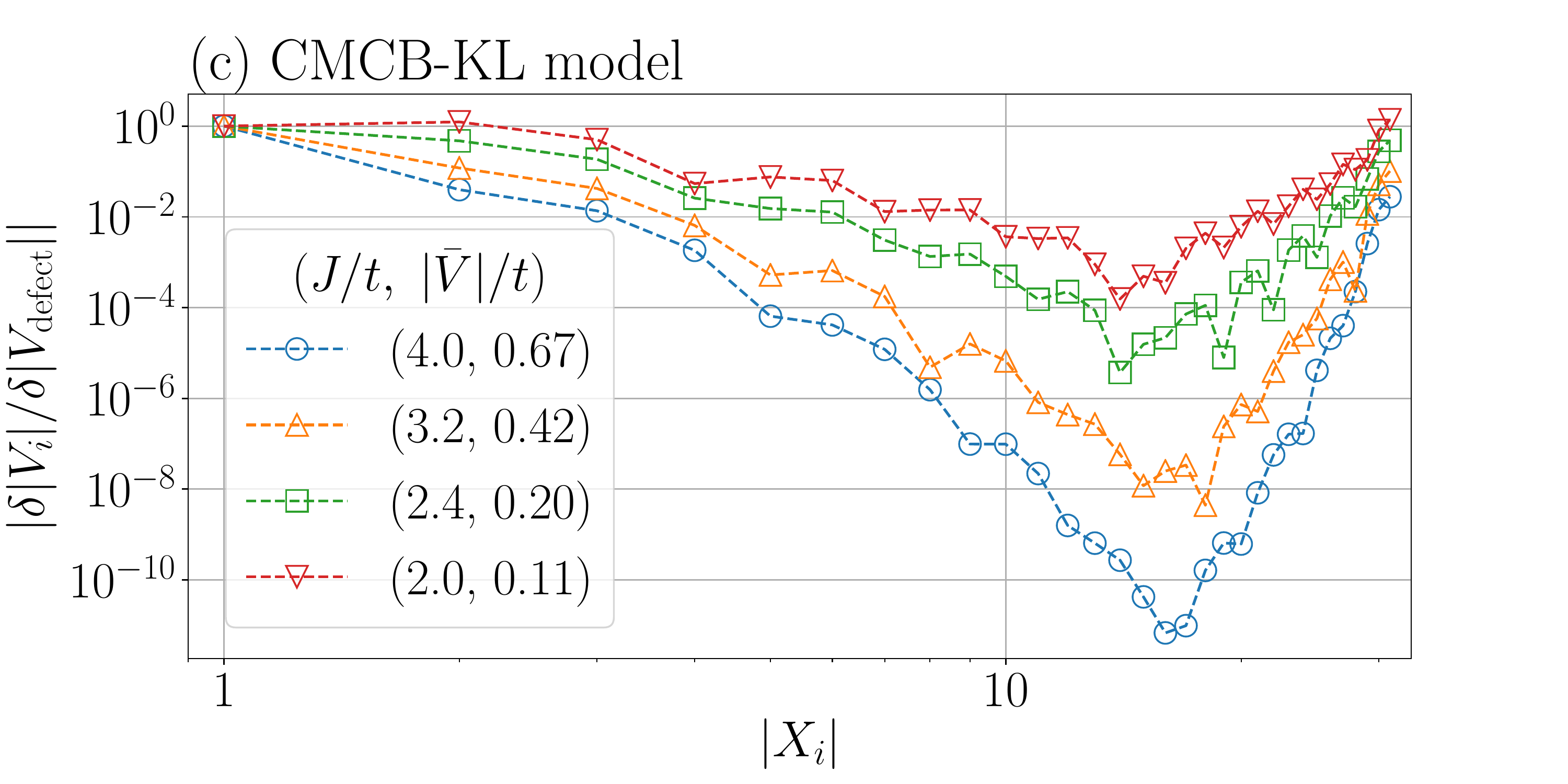}
   \\
   \vspace{-2mm}
    \includegraphics[width=93mm]{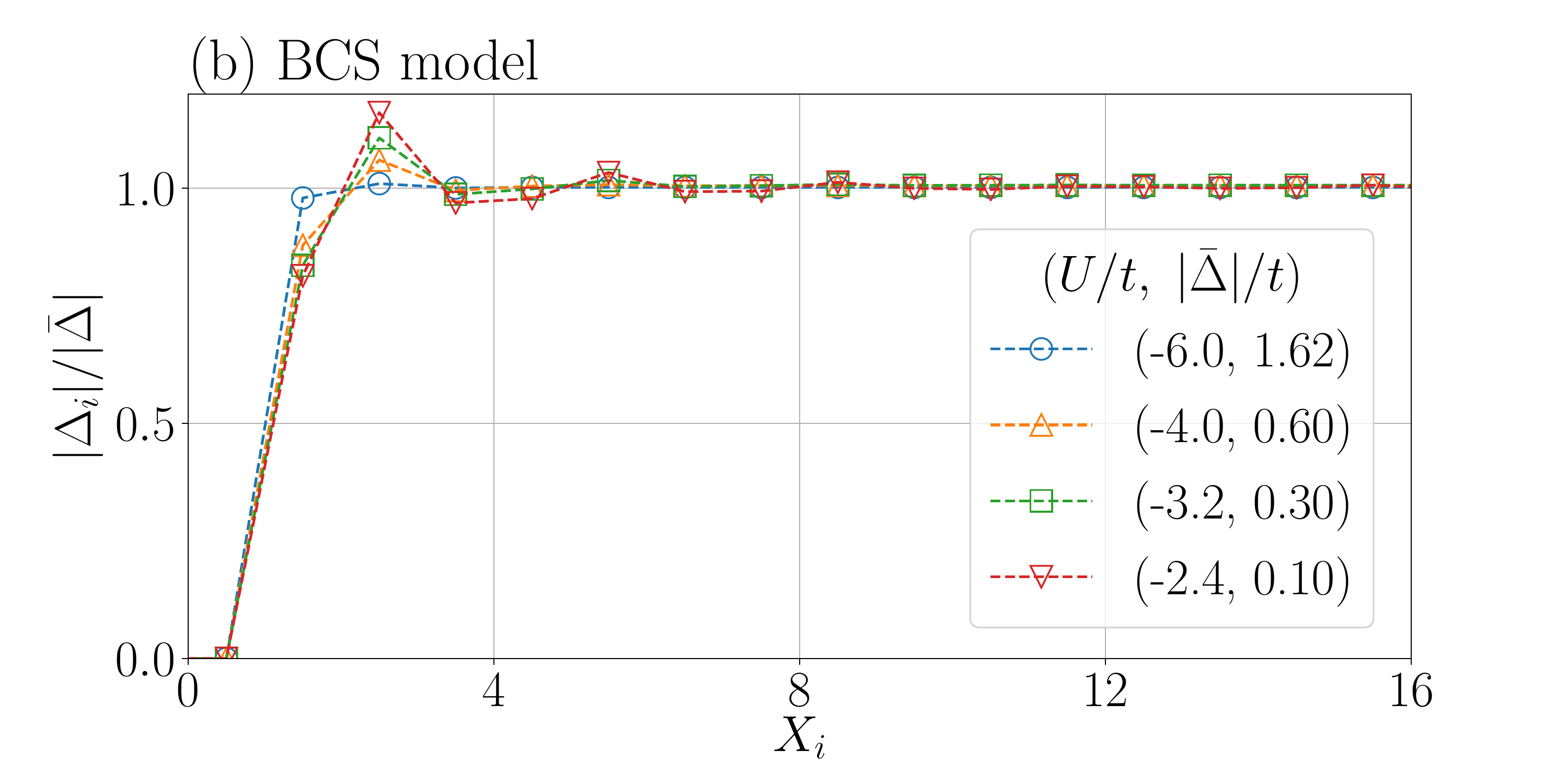}
    \hspace{-10mm}
    \includegraphics[width=93mm]{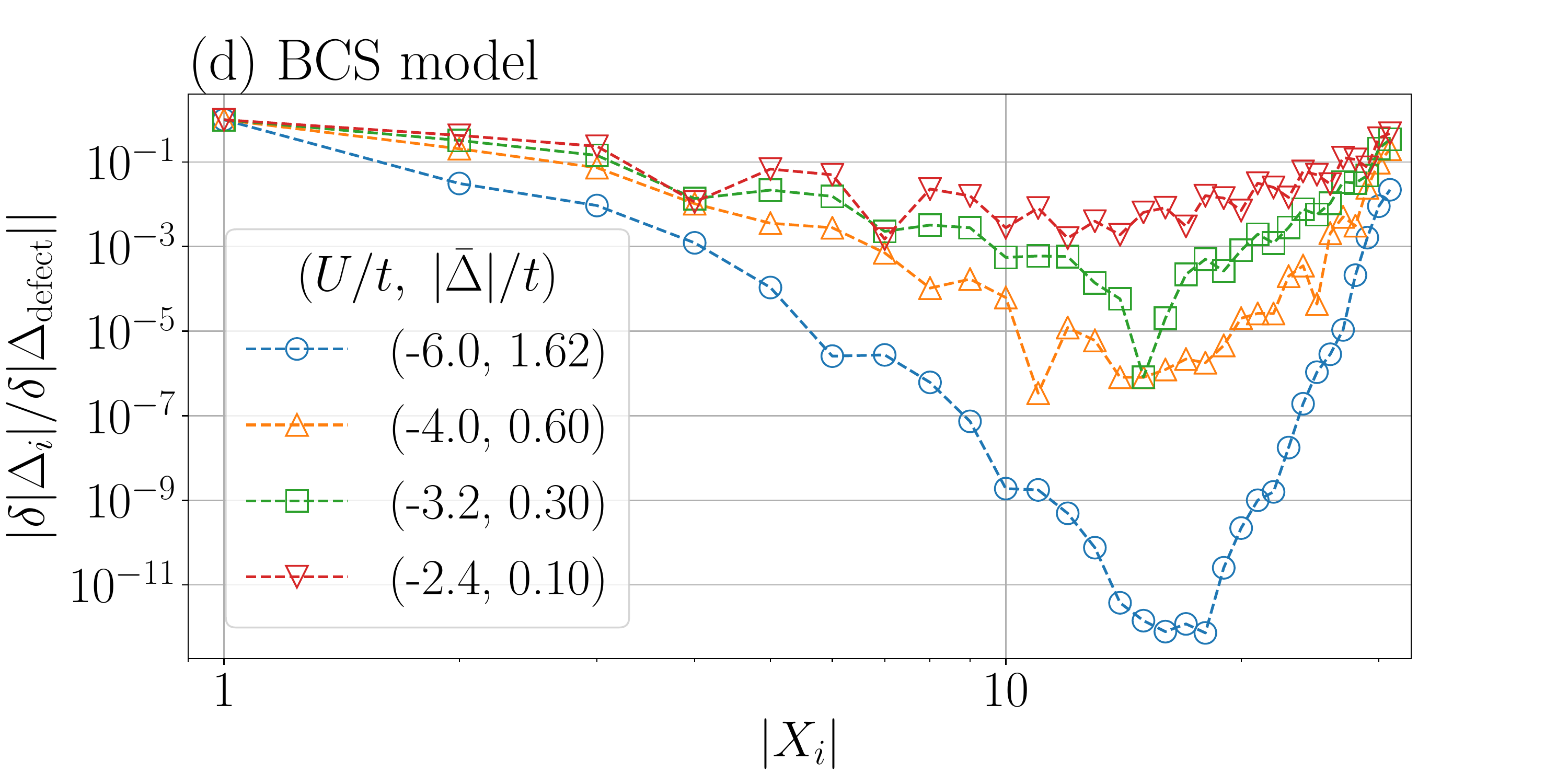}
  \caption{(Color online) Self-consistent solutions of the order parameters in the presence of the impurity potential. 
  Panels (a) and (b) respectively show the order parameters in the CMCB-KL and that in the BCS model. Panels (c) and (d) represent the derivative of the absolute value of the order parameters. Physical parameters are same as those in Fig.~\ref{fig:mean_fields_vortex}.}
  \label{fig:mean_fields_impurity}
\end{figure*}

We first study the characteristic length scale of the vortex. 
In the BCS superconductor, 
the parameter dependence of the length scale $\xi$ is given by $\xi \sim v_{\rm F}/\Delta$, where $\Delta$ is the bulk-gap of the superconducting state and $v_{\rm F}$ is the Fermi velocity. 
On the other hand, the parameter dependence of the length scale in the CMCB-KL is unclear because there are two-characteristic energy scales, i.e., the hybridization gap $|V|= |V_{\alpha \uparrow}|=|W_{{\bar \alpha}\downarrow}|$ and the Kondo-gap $\Delta_{\rm K} \sim \rho(0) |V|^2$, where $\rho(0)$ is the density of states at the Fermi level. 
Therefore we first look into the spatial variation of the order parameter to study the characteristic length scale in the CMCB-KL. 
The self-consistent solutions of the mean-field amplitudes are shown in Fig.~\ref{fig:mean_fields_vortex}. 
The absolute value of the mean-fields in the CMCB-KL is shown in the panel (a), where the condition $|V_{i1\uparrow}|=|V_{i2\uparrow}|=|W_{i1\downarrow}|=|W_{i2\downarrow}| \equiv |V_i|$ is always satisfied. 
The mean-field amplitude is slightly suppressed near the center of the vortex and is restored at a length of about the lattice constant. 
In addition, the spatial distribution of the mean-fields are well scaled by the real-space average of the order parameter $| {\bar V} |= \sum_i |V_i |/ N$. 
This parameter independent behavior is in contrast to the BCS superconductor where the core radius spreads out with decreasing the magnitude of the order parameter as shown in
Fig.~\ref{fig:mean_fields_vortex}(b) for comparison, where we show the order parameter of the BCS $s$-wave superconducting state obtained by solving the BdG equation for the attractive Hubbard model 
(See Appendix \ref{sec:BCS} for more details).
We see that the length scale becomes short even in the BCS theory when we use the strong attraction. 
Hence the appearance of the short length scale indicates that the superconducting electrons experience the large mean-field. 
As discussed later (Sec.\ref{sec:effective_theory}), this unusual property is connected to the characteristic pair potential  in the CMCB-KL.

To clearly show the parameter dependence of the length scale, we next calculate the derivative of the order parameters $\delta |V(X_i + 1/2)| = |V(X_i + 1)|-|V(X_i)|$ on discretized mesh of the tight-binding model. 
Figures~\ref{fig:mean_fields_vortex}(c) and \ref{fig:mean_fields_vortex}(d) respectively show the log-log 
plot of the derivative of the order parameter in the CMCB-KL and that in the BCS model.
In both of the figures, we see that the derivative shows the Friedel oscillation, whose periodicity is the order of $k_{\rm F}^{-1}$ where $k_{\rm F}$ is the Fermi wavenumber, and decays as it goes away from the center of the vortex ($X_i=0$) or from the system boundary ($X_i = 32$).

In Fig.~\ref{fig:mean_fields_vortex}(c) for the CMCB-KL, the length scale of the decay near the vortex core 
is not sensitive to the parameters. 
On the other hand, it is notable that the characteristic length scale near the boundary depends on the choice of the parameter and becomes larger with decreasing $|{\bar V}|$. 
Such a behavior is consistent with 
the coherence length 
$\xi= v_{\rm F}/|V|$ \cite{Iimura19}, where $v_{\rm F}$ and $|V|$ respectively denote the Fermi velocity and the mean-field amplitude in the homogeneous case. 
Hence the characteristic length scale for the vortex core is different from $\xi$. 
This fact implies the possibility that the characteristic length in the CMCB-KL is determined irrespective of the interaction term and hence is given by the lattice constant. 
In the BCS superconductor, for reference, both of the length scale near the vortex core and that near the boundary vary depending on the parameters 
as shown in Fig.~\ref{fig:mean_fields_vortex}(d).

To obtain further insight on the characteristic length scale in the CMCB-KL, 
we also calculate the self-consistent solutions in the presence of the non-topological defect instead of the vortex. 
We introduce the non-magnetic onsite potential $U_{imp}$ in the BdG Hamiltonian, which is given by
\begin{align}
U_{imp} = {\sum_{i \al}} u_i \left( n_{i\al\uparrow} + n_{i\al \downarrow} + n_{if\al} \right),
\end{align}
where $n_{if\al}=f_{i\al}^\dagger f_{i\al}$. 
The summation with respect to $i$ runs over $(X_i,Y_i) = \{ (\pm 1/2, 1/2) , (\pm 1/2,-1/2) \}$ around the origin. We take the amplitude of the potential as $u_i/t = 40$, so that the electrons and the pseudofermion cannot come to these impurity sites. 
In Fig.~\ref{fig:mean_fields_impurity}, we show the self-consistent solutions [Fig.~\ref{fig:mean_fields_impurity}(a) and Fig.~\ref{fig:mean_fields_impurity}(b)] and the derivative of the amplitude [Fig.~\ref{fig:mean_fields_impurity}(c) and Fig.~\ref{fig:mean_fields_impurity}(d)]. 
In the presence of the impurity potential, both of the order parameter in the CMCB-KL [Fig.~\ref{fig:mean_fields_impurity}(a)] and that in the BCS model [Fig.~\ref{fig:mean_fields_impurity}(b)] show similar behaviors. Hence, the characteristic length scale varies depending on the choice of the parameters, which is in contrast with the results in Fig.~\ref{fig:mean_fields_vortex}(b). 
Thus the parameter-independent behavior in the CMCB-KL is specific to the vortex state with nonzero winding number for the superconducting phase.

\subsubsection{Quasiparticle spectrum at vortex core}

\begin{figure}[t]
 \begin{center}
 \includegraphics[width=90mm]{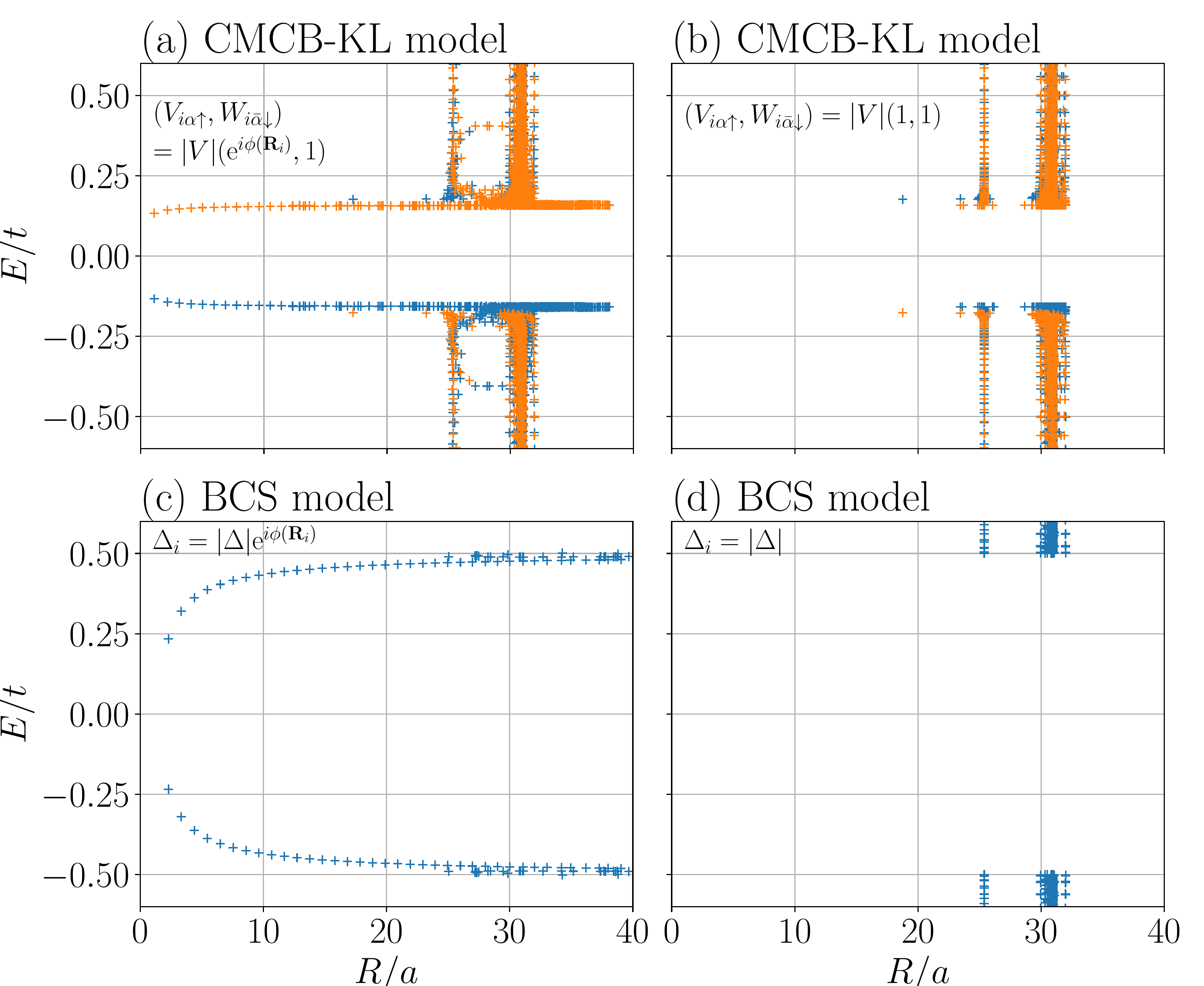}
  \caption{(Color online) Energy eigenvalue $\{ E_\gamma \}$ plotted as a function of $\{ R_\gamma \}$. Panels (a) and (b): The result in the CMCB-KL with $|V|/t=0.5$. 
  Left and right panel respectively show the result with $(V_{i\alpha \uparrow},W_{i{\bar \alpha}\downarrow}) = |V| (\mathrm{exp}[i \varphi({\bm R}_i)] , 1)$ (vorticity $\nu=1$) and the one with $(V_{i\alpha \uparrow},W_{i{\bar \alpha}\downarrow}) = |V| (1,1)$ ($\nu=0$). 
  Blue and orange markers respectively denote the contributions from ${\hat {\mathcal H}}_{\al=1}$ and ${\hat {\mathcal H}}_{\al=2}$. 
  Panels (c) and (d): The result in the BCS model with $|\Delta|/t=0.5$ is shown.  
  $E_{\rm F}/t=1$. $N_x=N_y=80$.}
  \label{fig:energy_dispersion}
 \end{center}
\end{figure}

Since we have confirmed that the very short length scale is a unique property of the vortex in the CMCB-KL, we expect the appearance of the unconventional quasiparticle excitations within the vortex core. 
Hence, we now examine the low-energy properties in the presence of the single vortex. 
We relate the spatial distribution of the wavefunction to the energy eigenvalues $\{ E^\al_{\gamma}\}$, which is obtained by diagonalizing the BdG Hamiltonian ${\hat {\mathcal H}}_\al$ ($\gamma=1,2,\cdots,3N$ is the index for eigenstates). 
For this purpose, we first introduce the quantity $R^\al_\gamma$ with the dimension of length as
\begin{align}
	\label{eq:CM}
	R^\al_{\gamma} = \sum_{I=(i,\nu)} \sqrt{ X_i^2 + Y_i^2} ~ |U^\al_{I\gamma}|^2,
\end{align}
which describes the real-space spread of the wavefunction.
Here $U^\al_{I\gamma}=({\hat U}^\al)_{I\gamma}$ denotes the eigenvectors of the BdG Hamiltonian.
The subscript $I~(=1,2,\cdots,3N)$ represents the index for both lattice $i$ and flavor $\zeta=( c_{\al \uparrow} , c^\dagger_{{\bar \al}\downarrow} ,f_\al )$ indices. 
For simplicity, we use the pair potential  given by $(V_{i\al\uparrow} ,W_{i{\bar \al}\downarrow})= |V| (\mathrm{e}^{i \nu \varphi({\bm R}_i)},1)$, where $\nu$ is the vorticity instead of solving the Eqs.\eqref{eq:v} and \eqref{eq:w} since the magnitude of the self-consistent solutions can be regarded as nearly constant in the CMCB-KL [See Fig.~\ref{fig:mean_fields_vortex}(a)].
This makes the calculation easier and the system size can be larger.

In Fig.~\ref{fig:energy_dispersion}, we show the relation between the energy eigenvalues $\{ E^\al_\gamma\}$ near the Fermi level and $\{ R^\al_\gamma \}$. 
Figures~\ref{fig:energy_dispersion}(a) and \ref{fig:energy_dispersion}(b) respectively show the result in the presence of the single vortex $( \nu=1)$ and that in the absence of the vortex ($ \nu=0$).
In Fig.~\ref{fig:energy_dispersion}(a), we see a characteristic behavior in $|E|/t \gtrsim0.2$, where $R_\gamma$ ($\sim 30$) seems to be independent on $E_\gamma$. 
Since such a behavior is also seen in the case without the vortex shown in Fig.~\ref{fig:energy_dispersion}(b), 
this behavior originates from the extended state in the homogeneous bulk case. 
Indeed, we can roughly estimate $R_\gamma \sim 0.4 \sqrt{N}$ when we assume the uniform solution $|U_{I\gamma}|^2 =1/(3N)$. 
This is consistent with $R_\gamma\sim30$ in this case with $N=80^2$.

On the other hand, the nearly flat branch, which continues to $R_\gamma \simeq 0$ appears in the low-energy region in the presence of the vortex [Fig.~\ref{fig:energy_dispersion}(a)]. 
This indicates that the energy eigenstates in the low-energy region is localized near the center of the vortex core. 
These characteristic behaviors are also seen in the similar plot for the BCS superconductor with the pair potential  $\Delta_i = |\Delta| \mathrm{e}^{i \nu \varphi({\bm R}_i)}$. The numerical results are respectively shown in Figs.\ref{fig:energy_dispersion}(c) ($\nu=1$) and \ref{fig:energy_dispersion}(d) ($\nu=0$).
The main difference between the CMCB-KL and the BCS model appears in the branches close to the vortex core, where the $R$-dependence is sharper in the BCS case. 

\begin{figure}[t]
 \begin{center}
    \includegraphics[width=88mm]{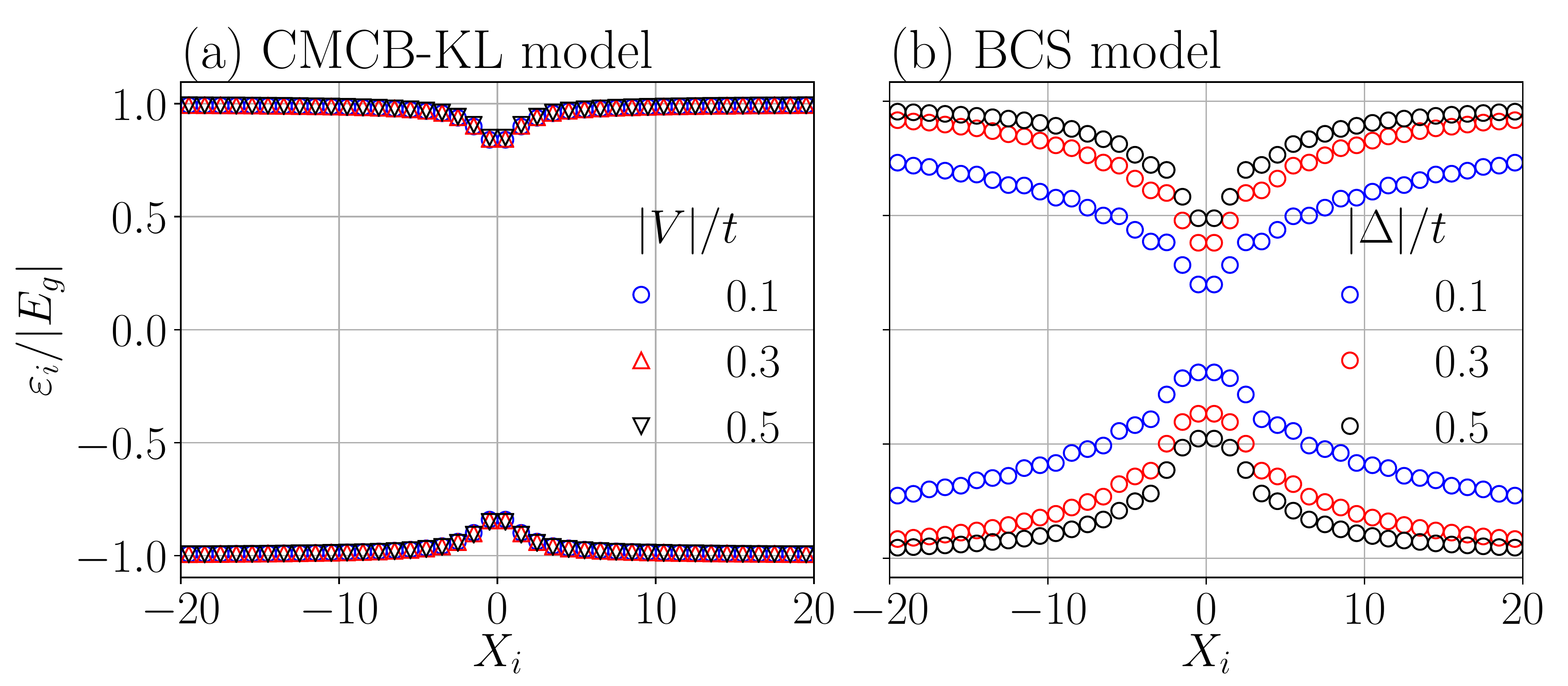}
  \caption{$\varepsilon_{i,\zeta=\uparrow}$ normalized by the magnitude of the bulk-gap $|E_g|$ is plotted as a function of $X_i$.
  In panel (a), blue-circle, red-upward triangle and black-downward triangle respectively correspond to $|V|/t=0.1$, $0.3$ and $0.5$. In panel (b), blue, red, and black markers correspond to $|\Delta|/t=0.1$, $0.3$, and $0.5$. 
  $N_x = N_y =80$.
  }
  \label{fig:energy_spectrum}
 \end{center}
\end{figure}

Since we have confirmed the localized nature of the eigenstates in the low-energy region inside the bulk gap, 
we next consider the energy distribution of the peak of the local density of states (DOS) 
$\rho_I (\omega) = \sum_{\gamma} |U_{I\gamma}|^2 \delta( \omega - E_\gamma )$ 
to elucidate the low-energy excitation of the quasiparticles within the vortex core. 
To this end, we 
define 
the local DOS in the low-energy region as  
\begin{align}
	F_{I}(\omega) = \frac{
\displaystyle
\sum_{\Lambda_- < E_\gm < \Lambda_+} |U_{I\gamma}|^2 
\delta( \omega - E_\gamma )
}
{
\displaystyle
\sum_{\Lambda_- < E_\gm < \Lambda_+} |U_{I\gamma}|^2 
}
\end{align}
where the distribution function $F_I(\omega)$ describes the energy profile of the local DOS in the low-energy region. 
We here impose the normalization condition $ \int_{\Lambda_-}^{\Lambda_+}d\omega ~F_I(\omega) = 1$ where the energy cutoff $\Lambda_\pm$, i.e., the bulk gap in the numerical calculation below is taken as the minimum of the energy gap defined in the absence of the vortex. 
The summation for $\gamma$ runs over the range where $\Lambda_-<E_\gamma<\Lambda_+$ is satisfied, to pick up the contributions from in-gap states. 

With the above information we can define the expectation value of the energy $\varepsilon_I$ as
\begin{align}
	& \varepsilon_I \equiv \int_{\Lambda_-}^{\Lambda_+}d\omega ~\omega F_I(\omega) = \frac{\displaystyle \sum_{\Lambda_-<E_\gamma<\Lambda_+} E_\gamma |U_{I\gamma}|^2}{\displaystyle \sum_{\Lambda_-<E_\gamma<\Lambda_+} |U_{I\gamma}|^2}.
\end{align}
$\varepsilon_I$ can be regarded as the peak position on the $\omega$-axis of the local DOS in $\Lambda_- < \omega < \Lambda_+$. 
In the following, we focus on the contribution from the conduction electron with $\sg=\uparrow$ since the BdG Hamiltonian holds the particle-hole symmetry $\rho_{i\uparrow}(\omega)=\rho_{i\downarrow}(-\omega)$. 

The numerical result in the CMCB-KL and that in the BCS model are respectively shown in Figs.\ref{fig:energy_spectrum}(a) and \ref{fig:energy_spectrum}(b), where the energy scale for each parameter is normalized by the magnitude of the minimum of the energy gap in the absence of the vortex (bulk gap $|E_g|$). 
In the CMCB-KL [Fig.~\ref{fig:energy_spectrum}(a)], the energy $\varepsilon_{i\uparrow}$ shows the full-gap behavior, which is well scaled by the magnitude of the bulk gap $|E_g|$ even in the limit $E_{\rm F}\gg |V|$. 
This is in strong contrast to the BCS model shown in Fig.~\ref{fig:energy_spectrum}(b), where the lowest excitation energy decreases with decreasing the magnitude of the pair potential  $|\Delta|$.  
This behavior is related to the fact that the minimal energy of the core state in usual $s$-wave superconductor is $\sim \Delta^2 /E_{\rm F}$.

Thus we have revealed that the vortex bound state in the CMCB-KL are characterized by the short length scale, which is independent of the parameter, and 
the quasiparticle energy at the vortex core is order of the bulk-gap.
We next 
clarify the physical origin of these characteristic properties which are contrast to the conventional full-gap 
BCS superconductor. 

\section{Low-energy effective theory}
\label{sec:effective_theory}
In this section, we construct the low-energy effective model in the continuum limit and the quasiclassical theory of the corresponding Green's function to study the physical origin of the characteristics of the low-energy quasiparticle in the CMCB-KL. 
Obviously, the short characteristic length scale with the order of the lattice constant is not compatible with the spirit of quasiclassical theory where we assume the presence of a long characteristic length.
However, we still have a possibility that the extrapolation of the quasiclassical theory works well also for the short-length scale range.
As demonstrated below, the quasiclassical theory for the CMCB-KL indeed qualitatively works, which is justified by comparing the results with those of tight-binding model discussed in the last section.

In Sec.\ref{subsec:model}, we introduce the effective Hamiltonian and the corresponding Green's function. 
In Sec.\ref{subsec:quasiclassical_theory}, we derive the Eilenberger equation and determine the normalization condition. 
Finally, we calculate the energy spectrum of the vortex bound state in the CMCB-KL analytically with the use of the Kramer-Pesch approximation \cite{Kramer74, Pesch74}, which was originally introduced to study the vortex bound state in the framework of the BCS theory.

\subsection{BdG Hamiltonian and Gor'kov equation in continuum limit}
\label{subsec:model}
\begin{figure}[t]
 \centering
    \includegraphics[width=88mm]{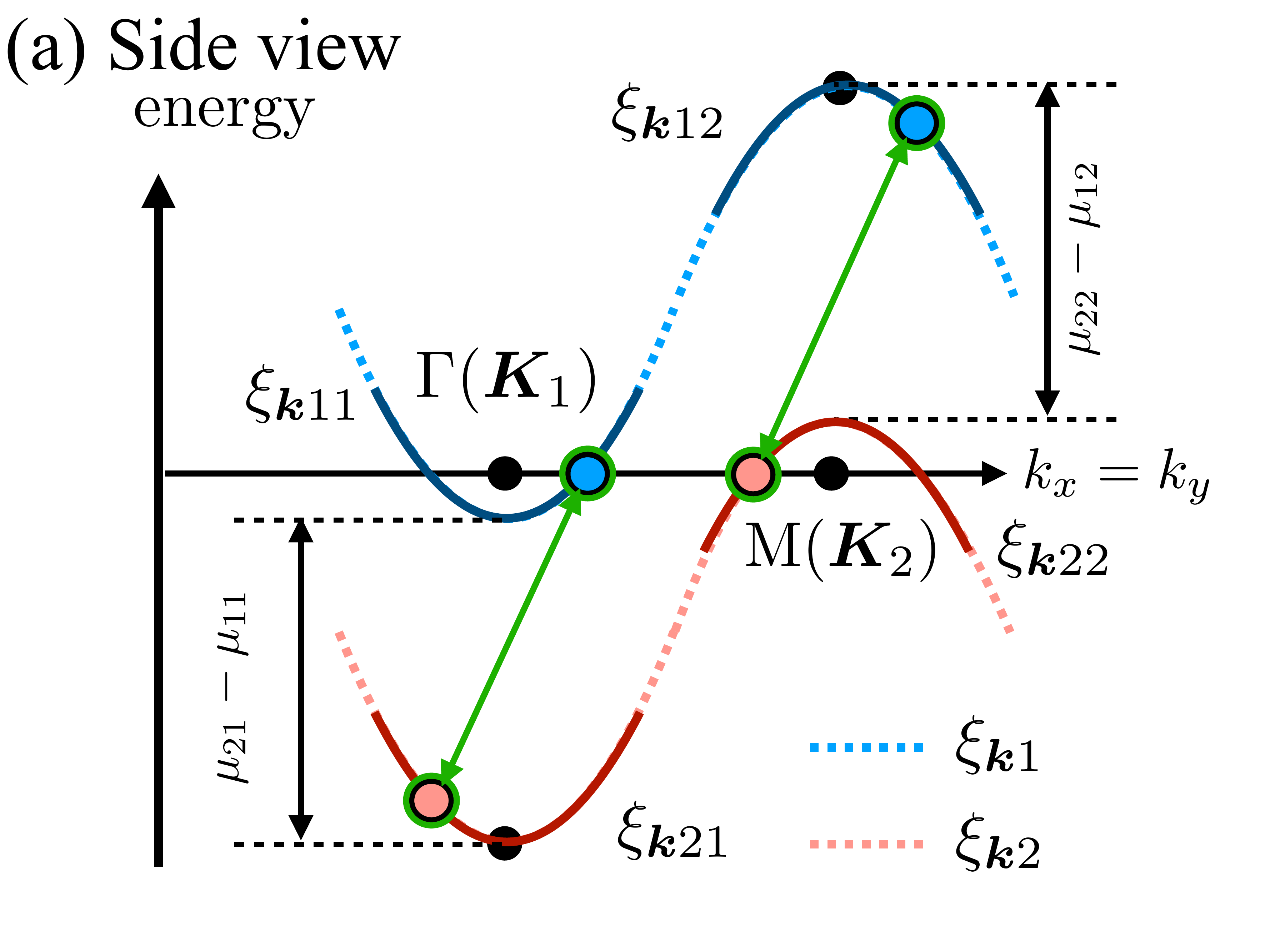}
    \\
    \vspace{-2mm}
    \includegraphics[width=88mm]{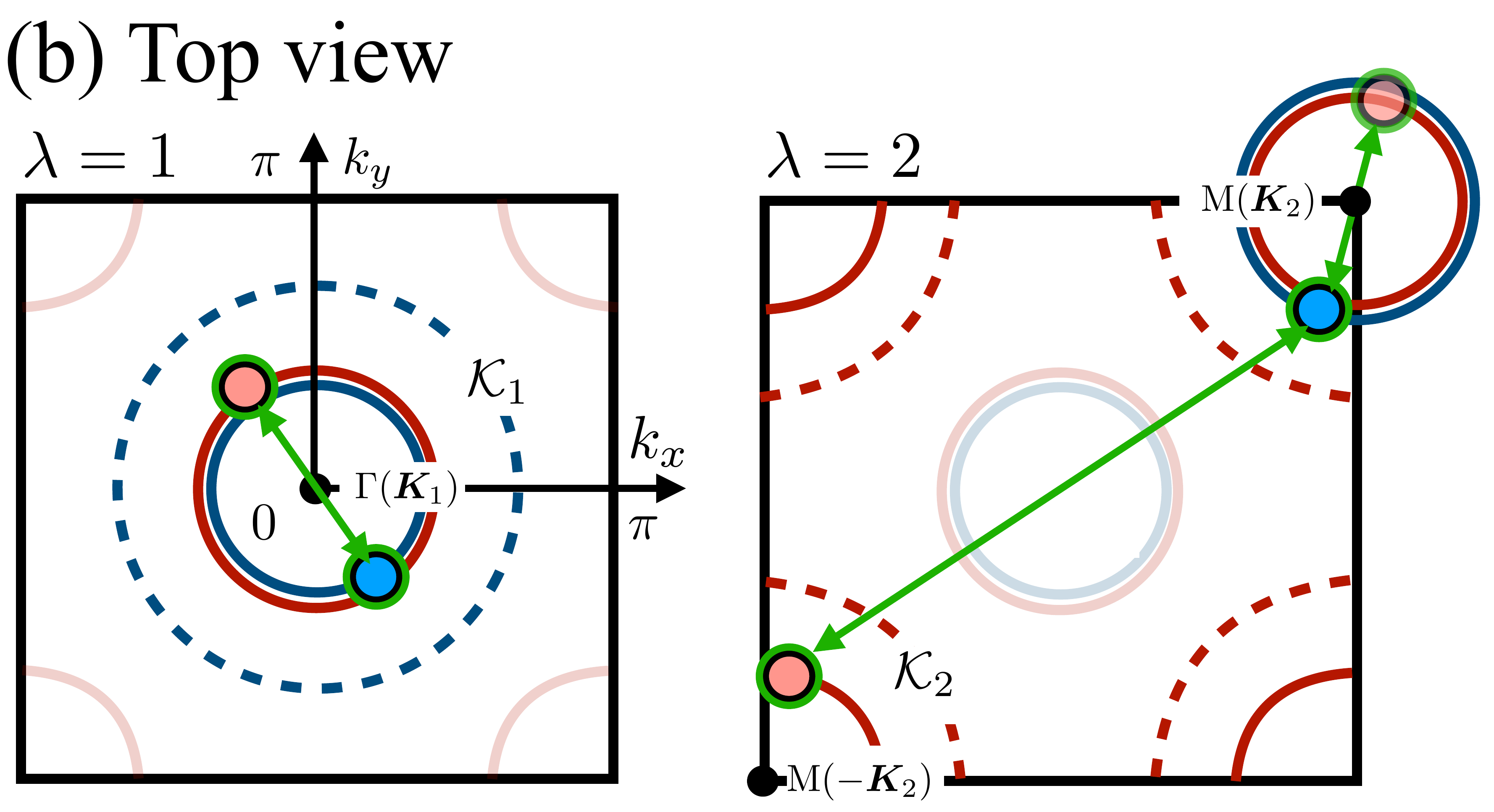}
  \caption{(Color online) Schematic picture of the effective model. Panels (a) and (b) respectively represent the side view and the top view of the dispersion relation. 
  Blue and pink dotted lines in panel (a) represent the original band dispersion given in Eq.\eqref{eq:band_dispersion}. 
  Parabolic dispersions defined in Eq.\eqref{eq:parabolic_dispersion} are drawn by red solid and navy solid lines. 
  The region ${\mathcal K}_\lambda$ is represented by dashed navy and red lines in panel (b).
  Blue and pink circle markers respectively denote the conduction electron in $\xi_{{\bm k}1}$ and that in $\xi_{{\bm k}2}$. 
  The Cooper pair formation is described by green arrows. 
  }
  \label{fig:schematic}
\end{figure}

In order to describe the non-uniform properties of the system, we construct the real-space representation of the CMCB-KL model with continuum approximation. 
To understand the physical origin of the numerical results obtained in the tight-binding model, we consider a simple model of the compensated metal obtained by approximating the one-particle kinetic energy of the electron band $\xi_{{\bm k}1}$ and the hole band $\xi_{{\bm k}2}$ with the parabolic dispersion. 
Assuming that the both Fermi surfaces are separated in the momentum space, we can approximate the energy dispersions as follows, 
\begin{align}
\label{eq:parabolic_dispersion}
\xi_{{\bm k}\al} &= \frac{\hbar^2}{2 m_{\al\lambda}} \left( {\bm k}- \sigma^z_{\al \al} {\bm K}_{\lambda} \right)^2 - \mu_{\al\lambda} ~({\bm k}\in {\mathcal K}_{\lambda}) \nonumber \\
&\equiv \xi_{{\bm k}\al \lambda},
\end{align}
where ${\bm K}_\lambda$ denotes the center of each Fermi surface. 
This model derives from the tight-binding model in Eq.\eqref{eq:band_dispersion} with the limit $\mu \to -4t$. 
We show the schematic of the parabolic dispersion in Fig.~\ref{fig:schematic}, where $\xi_{{\bm k}\al \al}~(\lambda=\alpha)$ and $\xi_{{\bm k}\al {\bar \al}}~(\lambda = {\bar \al})$ respectively describe the low-energy part of the $\alpha$-th band and the high-energy part. 
$m_{\al\lambda}$ and $\mu_{\al\lambda}$ respectively denote the effective mass and the chemical potential of the $\alpha$-th band in the region ${\mathcal K}_{\lambda}$ defined by $|{\bm k}-{\bm K}_{\lambda}|<k_c$, where $k_c$ denotes the band cut-off. 
In the BCS theory, the Cooper pairs are formed among the conduction electrons only in the low-energy region.
However, in our theory, not only the electron near the Fermi level, but also the ones in the high-energy region must be involved in the theoretical framework \cite{Iimura19}.
This is because the Fermi-surface-only model describes the composite pair amplitude given in Eq.\eqref{eq:composite}, but it cannot account for the Cooper pair amplitude composed of conduction electrons only \cite{Iimura19}.
Since the external field acts only on the conduction electrons, the presence of conduction electron pair is necessary for the electromagnetic properties including vortex state made of magnetic flux.
For this reason we need to consider the high-energy electron bands to produce Cooper pair of conduction electrons.
As shown below, 
our theoretical framework can reproduce the bulk properties. 

With the above parabolic dispersions,
we can obtain the real-space representation of the kinetic term as follows,
\begin{align}
& \sum_{{\bm k}\al\sg} c_{{\bm k}\al \sg}^\dagger \xi_{{\bm k}\al} c_{{\bm k}\al \sg} \nonumber \\
&\simeq \sum_{\al\lambda\sg} \int d{\bm r}~\psi_{\al\sg,\lambda}^\dagger ({\bm r}) \left[ - \frac{\hbar^2 {\bm \nabla}_{\al\lambda}^2}{2 m_{\al\lambda}}  - \mu_{\al\lambda} \right] \psi_{\al\sg,\lambda}^{ } ({\bm r}).
\end{align}
where the derivative operator is defined as ${\bm \nabla}_{\al\lambda} =  {\bm \nabla}-i \sigma^z_{\al \al} {\bm K}_{\lambda}$. 
$\psi_{\al \sg,\lambda}^{ } ({\bm r})$ is an annihilation operator of the conduction electron describing the degrees of freedom in the region ${\bm k}\in {\mathcal K}_{\lambda}$ and is defined as follows,
\begin{align}
	&\psi_{\al \sg}^{ } ({\bm r}) \simeq \sum_\lambda \psi_{\al \sg,\lambda}^{ } ({\bm r}),\\
	&\psi_{\al \sg,\lambda}^{ } ({\bm r}) = \frac{1}{(2\pi)^\frac{3}{2}} \int_{{\bm k}\in {\mathcal K}_\lambda} \hspace{-2em} d{\bm k} ~\psi_{{\bm k}\al\sg}^{ } \mathrm{e}^{i {\bm k}\cdot {\bm r}}.
\end{align}
Although we have taken the type-II limit in the numerical calculation, we introduce the vector potential to study the Meissner state. 
Following the gauge principle, the vector potential can be introduced as ${\bm \nabla} \psi_{\al\sg}^{ } ({\bm r}) \to ({\bm \nabla} - i e {\bm A}({\bm r}))\psi_{\al\sg}^{ } ({\bm r})$ (${\bm \nabla}\psi_{\al\sg}^\dagger ({\bm r}) \to ({\bm \nabla}+ie {\bm A}({\bm r}))\psi_{\al\sg}^\dagger ({\bm r})$) irrespective of the sign of the effective mass of the conduction band. 
This procedure is consistent with the result obtained by introducing the Peierse phase in the tight-binding model (see Appendix \ref{sec:current} for detail).

On the other hand, the real-space representation of the interaction term can be obtained by replacing the site index $i$ in Eq.\eqref{eq:hybridization_lattice} with the real-space coordinate ${\bm r}$. 
We then obtain the low-energy effective Hamiltonian in the presence of the magnetic field as follows,
\begin{align}
\label{eq:low_energy_effective_Hamiltonian}
&{\mathcal H}_{{\bm A}}^{\rm eff} = \sum_{\al \lambda} \int d{\bm r} \nonumber \\
&\hspace{0em}\times \left[ \psi_{\al\uparrow,\lambda}^\dagger ({\bm r}) \left( - \frac{\hbar^2 \grave{\bm D}_{\al \lambda}^2}{2m_{\al \lambda}} - \mu_{\al \lambda} \right) \psi_{\al\uparrow,\lambda}^{ } ({\bm r}) \right. \nonumber \\
& \hspace{0.5em}+ \left. \psi_{{\bar \al}\downarrow,\lambda}^{ } ({\bm r}) \left( \frac{\hbar^2 \acute{{\bm D}}_{{\bar \al}\lambda}^2}{2m_{{\bar \al}\lambda}} + \mu_{{\bar \al}\lambda} \right) \psi^\dagger_{{\bar \al}\downarrow,\lambda} ({\bm r}) \right. \nonumber \\
& \hspace{0.5em}+ \left. \left\{  \biggl( V_{\al \uparrow}^* ({\bm r}) \psi_{\al \uparrow,\lambda}^\dagger ({\bm r}) \right. \right. \nonumber \\
&\hspace{2.0em} \left. \left.  + W^*_{{\bar \al}\downarrow} \epsilon_{\al {\bar \al}} \psi_{{\bar \al}\downarrow,\lambda}^{ } ({\bm r})
\biggr) f_{\al \lambda}^{ } ({\bm r}) + \mathrm{H.c.} \right\} \right],
\end{align}
where $\grave{\bm D}_{\al \lambda} = {\bm \nabla}-i\sg^z_{\al\al} {\bm K}_{\lambda} - i e {\bm A}({\bm r})$ and $\acute{ {\bm D}}_{\al \lambda} = {\bm \nabla} +i\sg^z_{\al\al} {\bm K}_{\lambda}+ i e {\bm A}({\bm r})$. 
We have neglected the matrix elements of the interaction such as $ V_{1\uparrow} ({\bm r}) \psi_{1\uparrow,1}^\dagger ({\bm r}) f_{1,2}^{ } ({\bm r})$, which describe the scattering process with large momentum transfer, since we now focus on the quasiclassical limit $k_{\rm F}\xi \gg 1$, where $\xi$ is the characteristic length scale of the spatial non-uniformity of the mean-fields. 
Although the numerical results indicate that the characteristic length in the CMCB-KL is the atomic scale, in Sec.\ref{sec:Kramer_Pesch}, we show that our quasiclassical theory can be extrapolated to the quantum limit and reproduces the characteristics of the vortex bound state qualitatively. 

To study the low-energy bound state in the vortex core, we derive the Eilenberger equation, which is one of Green's function approaches used for spatially non-uniform superconductors \cite{Eilenberger68,Kopnin01}. 
Let us begin with the Gor'kov-Dyson's equation derived from the low-energy effective Hamiltonian in the Eq.\eqref{eq:low_energy_effective_Hamiltonian}. 
The one-particle Green's function is then defined as follows,
\begin{align}
&{\hat G}^c_{\al \lambda} (x,x^\prime) = \langle - T_\tau \left[ {\vec \psi}_{\al \lambda} (x) {\vec \psi}_{\al \lambda}^\dagger (x^\prime) \right] \rangle, 
\end{align}
where 
\begin{align}
{\vec \psi}_{\al \lambda}(x) = \mathrm{e}^{\tau {\mathcal H}^{\rm eff}_{{\bm A}}} ( \psi_{\al \uparrow,\lambda}^{ } ({\bm r}) ~\psi_{{\bar \al}\downarrow,\lambda}^\dagger ({\bm r}) )^T \mathrm{e}^{-\tau {\mathcal H}^{\rm eff}_{{\bm A}}}
\end{align}
and ${\vec \psi}^\dagger ({\bm r},\tau) = ( {\vec \psi}({\bm r},-\tau) )^\dagger$ are the Heisenberg representation of the Nambu basis. 
The corresponding Matsubara-Green's function is obtained as follows,
\begin{align}
{\hat G}^c_{\al \lambda} (i\omega_n;{\bm r},{\bm r}^\prime) &= \int_0^\beta d(\tau-\tau^\prime) ~\mathrm{e}^{i \omega_n (\tau-\tau^\prime)} {\hat G}^c_{\al \lambda} (x,x^\prime)\nonumber \\
&\equiv \left( 
\begin{array}{cc}
{\mathcal G}_{\al \lambda}(i\omega_n;{\bm r},{\bm r}^\prime) & {\mathcal F}_{\al \lambda}(i\omega_n;{\bm r},{\bm r}^\prime)\\
{\mathcal F}^\dagger_{\al \lambda}(i\omega_n;{\bm r},{\bm r}^\prime) & -{\bar {\mathcal G}}_{\al \lambda}(i\omega_n;{\bm r},{\bm r}^\prime)
\end{array}
\right).
\end{align} 
We have traced out the pseudofermions $f_{\al\lambda}(\bm r)$, which is an auxiliary degree of freedom introduced to describe the localized pseudospin moment. 
We then construct the self-contained theoretical framework involving only the conduction electrons for a given mean-field configuration.  

The Green's function obeys two-types of Dyson's equations, which are derived from the effective Hamiltonian as follows,
\begin{align}
\label{eq:left_hand_Gorkov}
&\delta^{(3)}({\bm r}-{\bm r}^\prime) {\hat \tau}^0\nonumber \\
&= \left( i \omega_n {\hat \tau}^0 - {\hat \xi}_{\al \lambda}^L ({\bm r}) - {\hat \Sigma}^c_{\al}(i\omega_n,{\bm r}) \right) {\hat G}_{\al \lambda}^c (i\omega_n ; {\bm r},{\bm r}^\prime), \\
\label{eq:right_hand_Gorkov}
&\delta^{(3)}({\bm r}-{\bm r}^\prime){\hat \tau}^0 \nonumber \\
&={\hat G}_{\al \lambda}^c (i\omega_n;{\bm r},{\bm r}^\prime) \left( i\omega_n {\hat \tau}^0 - {\hat  \xi}^R_{\al \lambda} ({\bm r}^\prime) -{\hat \Sigma}^c_{\al}(i\omega_n,{\bm r}^\prime) \right),
\end{align}
where ${\hat \tau}^0$ is the two-dimensional identity matrix. In addition, ${\hat \tau}^{i=1,2,3}$ used below is the Pauli matrix describing the degrees of freedom of the Nambu basis $\vec{\psi}_{\al\lambda}({\bm r})$. 
${\hat \xi}^{L}_{\al \lambda}({\bm r})$ denotes the kinetic energy, which is defined by
\begin{align}
\label{eq:kinetic_energy_operator}
&{\hat \xi}^L_{\al \lambda}({\bm r}) = \left( 
\begin{array}{cc}
\displaystyle -\frac{\grave{\bm D}_{\al \lambda}^2}{2 m _{\al \lambda}} -\mu_{\al \lambda} & 0  \\
0 & \displaystyle \frac{\acute{{\bm D}}_{{\bar \al}\lambda}^2}{2m_{{\bar \al}\lambda}} + \mu_{{\bar \al}\lambda} 
\end{array} 
\right),
\end{align}
and ${\hat \xi}^R_{\al \lambda}({\bm r})$ can be obtained by replacing $( {\grave {\bm D}}_{\al \lambda}, {\acute {\bm D}}_{{\bar \al} \lambda})$ with $({\acute {\bm D}}_{\al \lambda},{\grave {\bm D}}_{{\bar \al} \lambda})$. 
In addition, we have introduced the self-energy ${\hat \Sigma}^c_\al (i\omega_n;{\bm r})$, which is defined as follows
\begin{align}
\label{eq:self_energy_real_space}
{\hat \Sigma}^c_\al (i\omega_n;{\bm r}) &= \left(
\begin{array}{cc}
\Sigma_{\al\uparrow}(i\omega_n;{\bm r}) & \Delta_{\al} (i\omega_n;{\bm r}) \\
\Delta^\dagger_{\al} (i\omega_n;{\bm r}) & \Sigma_{{\bar \al} \downarrow}(i\omega_n;{\bm r})
\end{array}
\right),
\end{align}
with $\Delta_\alpha^\dg(i \omega_n;{\bm r}) =( \Delta_\alpha (-i \omega_n;{\bm r}))^* $, 
\begin{align}
& \Sigma_{\al\uparrow}(i\omega_n;{\bm r}) = \frac{|V_{\al\uparrow}({\bm r})|^2}{i\omega_n},\\
& \Sigma_{{\bar \al}\downarrow}(i\omega_n;{\bm r}) = \frac{|W_{{\bar \al}\downarrow}({\bm r})|^2}{i\omega_n},\\
\label{eq:anomalous_self_energy}
& \Delta_{\al} (i\omega_n;{\bm r}) = \frac{V_{\al\uparrow}^*({\bm r}) W_{{\bar \al}\downarrow}({\bm r}) \epsilon_{\al {\bar \al}} }{i\omega_n}.
\end{align}
We find that the anomalous self-energy (pair potential) $\Delta_\al$ is purely odd with respect to the fermionic Matsubara frequency $\omega_n$. 
Therefore, the superconductivity in the CMCB-KL is a new member of the odd-frequency superconductivity \cite{Berezinskii74,Balatsky92,Bergeret05,Tanaka12,Linder19}. 
In addition, the frequency dependence of the self-energy proportional to the inverse of the frequency implies that the pair potential is effectively enhanced in the low-energy region $\omega_n \to 0$.

\subsection{Quasiclassical Theory}
\label{subsec:quasiclassical_theory}
\subsubsection{Eilenberger equation}

We next derive the Eilenberger equation from the Dyson's equations \eqref{eq:left_hand_Gorkov} and \eqref{eq:right_hand_Gorkov}. 
We subtract Eq.\eqref{eq:right_hand_Gorkov} from Eq.\eqref{eq:left_hand_Gorkov} and expand the difference up to the first order of the spatial derivative ${\bm \nabla}_{{\bm r}_G}$, where ${\bm r}_G=({\bm r}+{\bm r}^\prime)/2$ is the center of mass coordinate of the two conduction electrons. 
Then, following the standard procedure \cite{Kopnin01}, 
we can integrate out the relative coordinate ${\bm R}={\bm r}-{\bm r}^\prime$ to obtain the Eilenberger equation, which is given by
\begin{align}
\label{eq:Eilenberger}
&\left[ {\hat B}_{\al \lambda} (i\omega_n,{\hat {\bm k}}_{{\rm F} \lambda}; {\bm r}_G ), {\hat g}_{\al \lambda} (i\omega_n,{\hat {\bm k}}_{{\rm F}\lambda} ; {\bm r}_G)  \right] \nonumber \\
&= i {\bm v}_{{\rm F}\lambda} \cdot {\bm \nabla}_{{\bm r}_G} {\hat g}_{\al \lambda} (i\omega_n,{\hat {\bm k}}_{{\rm F}\lambda} ; {\bm r}_G)
\end{align}
with 
\begin{align}
&{\hat B}_{\al \lambda} (i\omega_n,{\hat {\bm k}}_{{\rm F} \lambda}; {\bm r}_G ) \nonumber \\
&= {\hat \tau}^3 \left( i \omega_n - \frac{1}{2} E_c \sg^z_{\al \al} + e {\bm v}_{{\rm F}\lambda} \cdot {\bm A} ( {\bm r}_G)- {\hat \Sigma}_{\al}^c (i\omega_n;{\bm r}_G) \right) ,
\end{align}
where ${\hat {\bm k}}_{{\rm F}\lambda}={\bm k}_{{\rm F}\lambda} /|{\bm k}_{{\rm F}\lambda}|$ is the unit-vector of the Fermi momentum ${\bm k}_{{\rm F}\lambda}$, which is measured from the center of the Fermi surface ${\bm K}_\lambda$. 
${\bm v}_{{\rm F}\lambda} = {\bm k}_{{\rm F}\lambda}/m_\lambda$ is the Fermi velocity  
in the region ${\mathcal K}_\lambda$, where $m_\lambda=m_{1\lambda} =m_{2\lambda}$ is the effective mass of the conduction band in the region ${\mathcal K}_\lambda$. 
$E_c \equiv ( \mu_{\al \lambda}-\mu_{{\bar \al}\lambda}) \sg^z_{\al \al} > 0$ represents the band splitting of the conduction electrons (see Fig.\ref{fig:schematic}). 
${\hat g}_{\al \lambda} (i\omega_n,{\hat {\bm k}}_{{\rm F}\lambda} ; {\bm r}_G)$ is the quasi-classical Green's function, which is given by
\begin{align}
 \label{eq:quasi_classical_Green_function}
 &{\hat g}_{\al \lambda} (i\omega_n,{\hat {\bm k}}_{{\rm F}\lambda} ; {\bm r}) \nonumber \\
 &=\oint d\xi_{{\bm k}\lambda} \int d{\bm R}~{\hat G}_{\al \lambda}^c (i\omega_n;{\bm r}_+,{\bm r}_-) {\hat \tau}^3 \mathrm{e}^{-i {\bm k}\cdot {\bm R}} \nonumber \\ 
  &\equiv \left( \begin{array}{cc}
   g_{\al \lambda} (i\omega_n,{\hat {\bm k}}_{{\rm F}\lambda} ;{\bm r}) &- f_{\al \lambda} (i\omega_n,{\hat {\bm k}}_{{\rm F}\lambda} ;{\bm r}) \\
   f^\dagger_{\al \lambda} (i\omega_n,{\hat {\bm k}}_{{\rm F}\lambda} ;{\bm r}) & {\bar g}_{\al \lambda} (i\omega_n,{\hat {\bm k}}_{{\rm F}\lambda} ;{\bm r}) 
 \end{array} \right),
\end{align}
where ${\bm r}_\pm = {\bm r}\pm {\bm R}/2$. 
$\oint$ is an integration taking the contributions from the pole of the Green's function near the Fermi level $\xi_{{\bm k}\lambda \lambda} \equiv \xi_{{\bm k}\lambda}= 0$. 
In the following, we sometimes omit the argument $(i\omega_n,{\hat {\bm k}}_{{\rm F}\lambda};{\bm r})$ to simplify the presentation. 

To determine the non-uniform solution of the Eilenberger equation that is connected to the bulk state at the large enough distance,
we consider the normalization condition of the quasi-classical Green' s function. 
When ${\hat g}_{\al \lambda}$ satisfies Eq.\eqref{eq:Eilenberger}, ${\hat g}_{\al \lambda} {\hat g}_{\al \lambda}$ also becomes the solution of Eq.\eqref{eq:Eilenberger}. Hence, ${\hat g}_{\al \lambda}{\hat g}_{\al \lambda}$ can be written as
\begin{align}
	\label{eq:normalization_condition}
	{\hat g}_{\al \lambda} {\hat g}_{\al \lambda} = a {\hat \tau}^0 + b {\hat g}_{\al \lambda},
\end{align}
where ${\hat \tau}^0$ is the two-dimensional identity matrix and is the trivial solution of the Eilenberger equation. 
$a$ and $b$ are constants determined in a homogeneous case. 
The homogeneous solution can be derived from Eq.(\ref{eq:Eilenberger}). 
Here we need to care about the order of taking the limit.
Namely, we will take the band splitting $E_c$ as inifinity for the effective low-energy theory, but this must be done after performing the integrals.
Otherwise, we cannot pick up the leading-order contribution with respect to $E_c^{-1}$, which is necessary to form the conventional Cooper pair of conduction electrons \cite{Iimura19}. 
The resultant quasi-classical Green's function is given by
\begin{align}
	\label{eq:quasi_classical_Green_function_bulk}
	&{\hat g}_{\al \lambda} (i\omega_n) \nonumber \\
	&= \frac{\pi \mathrm{sgn}(\omega_n)\sg^z_{\al \al}}{\sqrt{\Omega_{n\al}^2 + \Delta_\al (i \omega_n) \Delta_\al^\dagger (i \omega_n)}} 
	\left( 
	\begin{array}{cc}
	i \Omega_{n\al} & -\Delta_\al (i\omega_n) \\
	\Delta_{\al}^\dagger (i\omega_n) & - i \Omega_{n\al} 
	\end{array}
	\right),
\end{align}
where $\Omega_{n\al}$ is defined as follows
\begin{align}
&i\Omega_{n\al}=i\omega_n - \frac{1}{2} E_c \sg^z_{\al \al} - \frac{1}{2} \left( \Sigma_{\al \uparrow}(i\omega_n) + \Sigma_{{\bar \al}\downarrow} (i\omega_n) \right).
\end{align}
Therefore, the normalization condition in Eq.\eqref{eq:normalization_condition} is obtained as follows
\begin{align}
& a = - \pi^2,\\
& b= 0.
\end{align}
These are identical to the normalization condition used in the BCS theory.

\subsubsection{Meissner response}
\label{sec:bulk_properties}
To test the derived Eilenberger equation, we calculate the charge current density, which is obtained in the quasi-classical theory as follows,
\begin{align}
	\label{eq:charge_current_density}
	{\bm j}=\frac{e}{\beta} \sum_{n \al} \rho_{\al}(0) \int \frac{d\Omega_{{\bm k}_{{\rm F}\al}}}{4\pi} {\bm v}_{{\rm F}\al} \mathrm{Tr} \left[ {\hat \tau}^3 {\hat g}_{\al \al}(i\omega_n,{\hat {\bm k}}_{{\rm F}\al}) \right],
\end{align}
where $d{\Omega}_{{\bm k}_{{\rm F}\al}}$ denotes the solid angle of the Fermi surface in ${\bm k}\in{\mathcal K}_\al$. 
See Appendix \ref{sec:current} for the derivation of the current density operator of the conduction electron with the effective mass. 
To calculate the linear response of the vector potential ${\bm A}({\bm r})$, we utilize the perturbative expansion of the quasi-classical Green's function given as ${\hat g}_{\al \lambda} = {\hat g}_{\al \lambda}^{(0)} + {\hat g}_{\al \lambda}^{(1)}$, where the overscript denote the order of the spatial derivative and the vector potential. The normalization condition in Eq.\eqref{eq:normalization_condition} gives,
\begin{align}
& g_{\al \lambda}^{(0)} + {\bar g}_{\al \lambda}^{(0)} =0,\\
& g_{\al \lambda}^{(1)} + {\bar g}_{\al \lambda}^{(1)} =0,\\
& ( g_{\al \lambda}^{(0)})^2 - { f}_{\al \lambda}^{(0)}  {f}^{\dagger(0)}_{\al \lambda}=-\pi^2,\\
& 2 g^{(1)}_{\al \lambda} g^{(0)}_{\al \lambda} =f_{\al \lambda}^{(1)} f^{\dagger (0)}_{\al \lambda} + f_{\al \lambda}^{(0)} f^{\dagger (1)}_{\al \lambda}.
\end{align}
The zeroth-order solutions of the quasi-classical Green's functions are same as Eq.\eqref{eq:quasi_classical_Green_function_bulk}. 
With the use of the normalization condition, we can obtain the two-independent Eilenberger equations describing the first-order corrections, which are given by
\begin{align}
	&\Delta_\al g_{\al \lambda}^{(1)} - i \Omega_{n\al} f_{\al \lambda}^{(1)} = \frac{1}{2} i {\bm v}_{{\rm F}\lambda} \cdot \left( {\bm \nabla}-2ie {\bm A}({\bm r}) \right) f_{\al \lambda}^{(0)},\\
	&i \Omega_{n\al} g_{\al \lambda}^{(1)} - \Delta_\al^\dagger f_{\al \lambda}^{(1)} = \frac{1}{2} i {\bm v}_{{\rm F}\lambda} \cdot {\bm \nabla} g_{\al \lambda}^{(0)}.
\end{align}
From above, the first order term is obtained as follows
\begin{align}
	&g_{\al \lambda}^{(1)} \nonumber\\
	&= - \frac{ \pi \mathrm{sgn}(\omega_n) \sg^z_{\al \al} |V_{\al \uparrow}^* W_{{\bar \al}\downarrow}|^2 }{\omega_n^2 (\Omega_{n\al}^2+\Delta_\al \Delta_\al^\dagger)^\frac{3}{2} }  e {\bm v}_{{\rm F}\lambda} \cdot \left( {\bm A}({\bm r}) - \frac{1}{2e} {\bm \nabla} \theta({\bm r}) \right).
\end{align}
where we have assumed that $\Delta_\al (i\omega_n) = |V_{\al \uparrow}^* ({\bm r}) W_{{\bar \al}\downarrow} ({\bm r}) | \mathrm{e}^{i \theta ({\bm r})} \epsilon_{\al {\bar \al}} /(i\omega_n)$. 
The resultant quasiclassical Green's function is invariant for the gauge transformation ${\bm A} ({\bm r})\to {\bm A} ({\bm r})+ {\bm \nabla} \chi ({\bm r})$ with $\theta({\bm r}) \to \theta ({\bm r})+ 2e \chi ({\bm r})$. 
This fact supports that the physical U(1) gauge degrees of freedom is only the relative phase $\theta({\bm r})$ between the conduction electron $\psi_{\al \uparrow}$ and $\psi_{{\bar \al}\downarrow}$. 
When we consider the Meissner state, the U(1) gauge $\theta({\bm r})$ is fixed.  
Then the Fourier component of the charge current density is given by ${\bm j}({\bm q}) = - K({\bm q}) {\bm A}({\bm q})$. 
From Eq.\eqref{eq:charge_current_density}, we can obtain the Meissner kernel $K({\bm q}\to {\bm 0})\equiv K$, which is given by
\begin{align}
	&K\equiv \sum_{\al} \frac{n^{SC}_\al (T) e^2}{|m_\al|},
\end{align}
where $n_\al^{SC}(T)$ is the superfluid density, which is obtained as follows,
\begin{align}
\label{eq:superfluid_density}
	&n_\al^{SC} (T) = n_\al  \frac{1}{\beta}\sum_n \frac{\pi \mathrm{sgn}(\omega_n) \sg^z_{\al \al} |V^*_{\al \uparrow} W_{{\bar \al} \downarrow}|^2}{\omega_n^2 (\Omega_{n\al}^2+\Delta_\al \Delta_\al^\dagger)^\frac{3}{2} }.
\end{align}
$n_\al=4 \rho_\al (0) |\mu_{\al \al}|/3$ is the number density of the conduction electron ($\alpha =1$) or that of the hole ($\alpha=2$). 
In the low-temperature limit, the fermionic Matsubara-frequency $\omega_n = (2n+1) \pi T$ can be regarded as a continuous valuable. 
We hence consider the partitioning quadrature, which is defined as
\begin{align}
	\lim_{T\to 0} T \sum_n F(\omega_n) = \frac{1}{2\pi} \int_{-\infty}^\infty d\omega~F(\omega).
\end{align}
Then, we can rewrite the superfluid density as follows
\begin{align}
	& \frac{n_\al^{SC}(T\to 0)}{n_\al} \simeq \left( \frac{|V|}{E_c} \right)^2 \int_{-\infty}^\infty dx  \frac{-4 i \mathrm{sgn}(x)  }{x^2 \left[1 - i \frac{4}{x}\right]^\frac{3}{2}}  = \frac{4 |V|^2}{E_c^2},
\end{align}
where we have used $|V_{\al \uparrow}|=|W_{{\bar \al}\downarrow}|=|V|$ and neglected the higher-order contributions of the order of $|V|/E_c$. 
This result is confirmed also by carrying out the summation with respect to the Matsubara frequency numerically.
From above, we can obtain the magnetic penetration depth $\lambda$ as follows,
\begin{align}
	\label{eq:magnetic_penetration_depth}
	\lambda= \sqrt{\frac{1}{\mu_0 K}} = \frac{E_c}{2|V|} \left( \mu_0 \sum_\al \frac{n_\al e^2}{|m_\al |} \right)^{-\frac{1}{2}}.
\end{align}
The result is same as the one derived by calculating the current-current correlation function without the quasi-classical approximation \cite{Iimura19}. 
Since the resultant magnetic penetration depth is larger than the typical value in the BCS theory by the order of $E_c/|V| \gg 1$, the superconducting state of the CMCB-KL can be regarded as an extreme limit of the type-II superconductor. 

In the above derivation, we have learned a lesson relevant to our CMCB-KL:
when we focus on the low-temperature properties, 
we cannot take the limit $E_c\to \infty$. 
This fact is symbolically expressed by 
\begin{align}
\left[ \lim_{E_c \to \infty} ,~\lim_{T\to 0} T \sum_{n=-\infty}^\infty \right] \neq 0.
\end{align}
This is associated with the fact that the self-energy is enhanced in the low-energy region due to the $1/i \omega_n$ shape, and it can become larger than the band splitting $E_c$ at low temperatures.

\subsection{Application of Kramer-Pesch approximation and vortex bound state in CMCB-KL}
\label{sec:Kramer_Pesch}
\subsubsection{Binding energy of vortex core state}
To elucidate the low-energy properties of the quasiparticle excitation in the vortex-core state, we use the perturbative theory introduced by Kramer-Pesch \cite{Kramer74, Pesch74}. 
Since the magnetic penetration depth in Eq.\eqref{eq:magnetic_penetration_depth} is larger than the usual value in the BCS theory, we consider the type II limit to neglect the electromagnetic field ${\bm A}$ in the following discussion. 
We now assume that the isolated vortex, which is described by the following pair potential with single flux quanta: 
\begin{align}
	\Delta_\al (z, {\bm r}) = \frac{|V ({\bm r})|^2 \mathrm{e}^{i \varphi}}{z} \epsilon_{\al {\bar \al}},
\end{align}
where $\varphi$ is the azimuth angle of the two-dimensional polar coordinate system. 
$z$ is the complex frequency.
We set $|V({\bm r})|=|V_{\al \uparrow}({\bm r})|=|W_{{\bar \al}\downarrow} ({\bm r})|$ as confirmed in the tight-binding model simulation in Sec.\ref{sec:tight_binding_model}. 
We also take $|V({\bm r})| = |V(r)|$, where $r=|{\bm r}|$ since we have assumed the $s$-wave symmetry of the superconducting state. 
To write the Eilenberger equation in the simplest form, it is better to use another coordinate system, where the axis is parallel to the Fermi velocity \cite{Kopnin01}. 
Then, the position of the quasiparticle is specified by three parameters; the direction of the Fermi velocity being at angle $\gamma$ with the $x$-axis, 
the impact parameter $b$ measured from the vortex core and the distance $u$ along the quasiparticle trajectory. 
In order to calculate the bound state energy, we can put $\gamma=0$ since the Fermi surface is isotropic in our effective model. 
Then, the Eilenberger equations are written as follows,
\begin{align}
\label{eq:Eilenberger_eq_f_KP}
&- i v_{{\rm F}\lambda} \partial_u f_{\al \lambda} = 2\Lambda_{\al}(z) f_{\al \lambda} - 2 \Delta_\al (z) g_{\al \lambda},\\
\label{eq:Eilenberger_eq_fd_KP}
&- i v_{{\rm F}\lambda} \partial_u f^\dagger_{\al \lambda} = -2 \Lambda_{\al}(z) f^\dagger_{\al \lambda} + 2 \Delta^\dagger_\al (z) g_{\al \lambda},
\end{align}
where $v_{{\rm F}\lambda}$ is the magnitude of the Fermi velocity. $\Lambda_\al (z)$ is defined as follows,
\begin{align}
	&\Lambda_\al (z) = z - \frac{1}{2} E_c \sg^z_{\al \al} - \frac{1}{2} \left( \Sigma_{\al \uparrow}(z) + \Sigma_{{\bar \al}\downarrow}(z) \right).
\end{align}
We note the relation $\Lambda_\al(i\omega_n) = i \Omega_{n\al}$. 
To solve the Eilenberger equation with the use of the perturbative approach, we first focus on the boundary condition of the vortex bound state. 
In the literature which applies the Kramer-Pesch approximation for the conventional BCS superconductor \cite{Kramer74,Pesch74}, 
the low-energy excitation appears in the region $|\omega| \ll |\Delta_{\rm BCS}|$ and the frequency $\omega \simeq 0$ has been treated as the perturbation. 
On the other hand, we need to reconsider the energy region where the perturbative approach is justified in the CMCB-KL since the vortex bound state has a characteristic energy, which is order of the Kondo-gap $|\omega| \sim |V|^2/E_c$ rather than $\omega\simeq0$.
[See Fig.~\ref{fig:energy_spectrum}(a).]
Hence we consider the boundary condition for $g_{\al \lambda}$. 
In the bulk limit $|u|\to \infty$, the quasiclassical Green's function $g_{\al \lambda}$ asymptotically approaches to
\begin{align}
	\label{eq:eq:boundary_condition_g}
	&g^{R,A}_{\al \lambda} (\omega,|u|= \infty) \nonumber \\
	&=  (-1)^{R,A} i \pi \sg^z_{\al \al} \frac{\Lambda_\al(\omega \pm i \delta)}{\sqrt{ \Lambda_\al^2(\omega \pm i\delta) - \frac{|V_\infty |^4}{(\omega\pm i\delta)^2} }},
\end{align}
where $|V_\infty| = |V_{\al \uparrow}(\pm \infty)| = |W_{{\bar \al}\downarrow}(\pm \infty)|$ represents the amplitude of the order parameter in the bulk limit. 
The overscripts $R$ and $A$ respectively stand for the retarded- and the advanced version of the quasi-classical Green's function. 
$\delta >0$ is a positive infinitesimal. 
The sign $(-1)^{R,A}$ which is $(+1)$ for $R$ and $(-1)$ for $A$ results from the analytical continuation. 

On the other hand, 
as shown in Fig.~\ref{fig:energy_spectrum}(a), the peak-position of the local DOS in the low-energy region is localized near the center of the vortex. 
This indicates that the quasiclassical Green's function $g_{\al \lambda} (\omega_0,u)$, where $\omega_0$ represents the binding energy of the low-energy quasiparticle must vanish in the bulk limit $|u|\to \infty$. 
This fact requires $\Lambda_\al (\omega_0,|u|=\infty)=0$ so that $g_{\al \lambda} (\omega_0 , |u| =\infty)=0$. 
Hence, we can obtain the equation relating to $\omega_0$ as follows,
\begin{align}
&\Lambda_{\al} (\omega_0,|u|=\infty)= \omega_0 - \frac{1}{2} E_c \sg^z_{\al \al}
- \frac{|V_\infty|^2}{\omega_0} = 0.
\label{eq:omega0_eq}
\end{align}
From above, we can find two-solutions of $\omega_0$, which are obtained as follows,
\begin{align}
\omega_0^{\eta} &= \frac{1}{4} \left[ E_c \sg^z_{\al \al} + \eta \sqrt{E_c^2 + 16 |V_\infty|^2 }~\right] ~(\eta=\pm 1).
\end{align}
The solution for $\eta=-\sg^z_{\al \al}$ is order of the Kondo-gap, while the other one is far away from the Fermi level. 
Hence, we regard the former one as the binding energy of the vortex bound state $\varepsilon_{\al0}$, which is obtained as follows,
\begin{align}
	&\varepsilon_{\al 0} \simeq - \sg^z_{\al \al} \frac{2 |V_\infty|^2}{E_c}.
\end{align} 
If we rewrite the energy splitting $E_c$ in terms of the potential in the tight-binding model as $E_c = 2|\mu|$, 
we obtain $\varepsilon_{\al 0} \simeq - \sg^z_{\al \al} |V_\infty|^2/|\mu|$, which is same as the Kondo-gap evaluated in the low-energy effective theory \cite{Iimura19}.
In addition, this binding energy is also consistent with the numerical result shown in Fig.~\ref{fig:energy_dispersion}(a), where the electron band ($\alpha=1$) forms the localized state with the negative energy, while the one with the positive energy is composed of the hole band ($\alpha=2$). 
Therefore, the characteristic energy scale of the vortex bound state in the CMCB-KL results from characteristic frequency dependence of the $normal$ self-energies $\Sigma_{\al \uparrow}(\omega)$ and $\Sigma_{{\bar \al}\downarrow}(\omega)$ as reflected in the third term in the middle of Eq.~\eqref{eq:omega0_eq}. 
If we apply the above scheme for the BCS superconductors, we get $\omega_0 = 0$ \cite{Kopnin01} since $E_c$ and $|V_\infty|^2/\omega_0$ terms are absent in Eq.~\eqref{eq:omega0_eq} are absent.

\subsubsection{Characteristic length scale and energy dispersion}
So far, we have focused on the boundary condition in the bulk limit $|u|=\infty$ with $b=0$ to determine the binding energy of the vortex core state. 
We now solve the Eilenberger equations \eqref{eq:Eilenberger_eq_f_KP} and \eqref{eq:Eilenberger_eq_fd_KP} since the spatial derivative with respect to the coordinate $u$ includes the information of the characteristic length scale of the core state. 
From Eqs.\eqref{eq:Eilenberger_eq_f_KP} and \eqref{eq:Eilenberger_eq_fd_KP}, we find the following symmetries
\begin{align}
	f_{\al \lambda} (u)  &= - f_{\al \lambda}^\dagger (-u),\\
	g_{\al \lambda} (u) &= g_{\al \lambda} (-u).
\end{align}
Hence, we focus on the Eq.\eqref{eq:Eilenberger_eq_f_KP} in the following. 
Assuming $u \gg b \gtrsim 0$ and $z\simeq \ep_{\al 0}$, we regard $\Lambda_\al (z) \propto ( z - \varepsilon_{\al0} )$ and the impact parameter $b$ as the perturbation. 
In addition, the pair potential is rewritten in the $(u,b)$ coordinate system as follows, 
\begin{align}
	\label{eq:pair_potential_KP}
	\Delta_\al (z,u,b) &= \frac{|V(u,b)|^2}{z} \epsilon_{\al {\bar \al}} \frac{u + i b}{\sqrt{u^2 + b^2}} \nonumber \\
	&= {\bar \Delta_\al}(z,u) \mathrm{sgn}(u) \left( 1 + i \frac{b}{u} \right) + \cdots,
\end{align}
where ${\bar \Delta}_\al (z,u) = |V(u)|^2 \epsilon_{\al{\bar \al}} /z$ describes the frequency dependence of the anomalous self-energy. 
Then, the quasi-classical Green's functions up to the first-order of the perturbation obey the following equations
\begin{align}
	&-i v_{{\rm F}\lambda} \partial_u f_{\al \lambda}^{(0)} (u) = -2 {\bar \Delta}_\al (z,u) \mathrm{sgn}(u) g_{\al \lambda}^{(0)} (u), \\
	&-i v_{{\rm F}\lambda} \partial_u f_{\al \lambda}^{(1)} (u) \nonumber \\
	&\hspace{1em} = 2 \Lambda_\al (z,u) f_{\al \lambda}^{(0)}(u) - 2 i {\bar \Delta}_\al (z,u) \frac{b}{|u|} g_{\al \lambda}^{(0)} (u),
\end{align}
where the overscript represents the order of the perturbation. 
We here assume that $f_{\al \lambda}^{(0)} (u)= - i  g_{\al \lambda}^{(0)}(u)$ to make $g_{\al \lambda}$ satisfy the boundary condition $g_{\al \lambda}(\pm \infty)=0$. 
We then obtain the quasi-classical Green's function as follows,
\begin{align}
	\label{eq:quasiclassical_g_KP}
	&g_{\al \lambda}(u) = C_0 \mathrm{e}^{- K_{\al \lambda} (z,u)}, \\
	\label{eq:quasiclassical_f_KP}
	&f_{\al \lambda}(u) =- i \mathrm{e}^{i \gamma} C_0 \biggl[ \mathrm{e}^{- K_{\al \lambda}(z,u)}  \biggr. \nonumber \\
	&\left. + \frac{2 i }{v_{{\rm F}\lambda}} \int_0^u du^\prime \left(\Lambda_\al (z,u^\prime) +  {\bar \Delta}_\al (z,u^\prime) \frac{b}{|u^\prime|} \right) \mathrm{e}^{-K_{\al \lambda}(z,u^\prime)}\right]
\end{align}
where $C_0$ is the coefficient, which is determined below so that $f_{\al \lambda}(u)$ satisfies the boundary condition in the bulk limit. 
$K_{\al \lambda}(u)$ is defined as follows,
\begin{align}
	K_{\al \lambda} (z,u) &=- \frac{2}{v_{{\rm F}\lambda}} \int^{|u|}_0 du^\prime  {\bar \Delta}_{\al} (z,u^\prime) \nonumber \\
	&\simeq \frac{E_c}{v_{{\rm F}\lambda}} \int_0^{|u|}du^\prime \frac{|V(u^\prime)|^2}{|V_\infty|^2}.
\end{align}
Since the amplitude of the order parameters can be regarded as nearly constant over the whole space as evidenced by the tight-binding model calculation [See Fig.~\ref{fig:mean_fields_vortex}(a)], we can find the characteristic length scale of the vortex bound state ${\tilde \xi}$ from $g_{\al\lambda}(u) \sim e^{-u/\tilde \xi}$ as follows,
\begin{align}
	{\tilde \xi} \equiv \frac{v_{{\rm F}\lambda}}{2 | {\bar \Delta_\al}(\varepsilon_{\al 0},\infty)|} \simeq \frac{v_{{\rm F}\lambda}}{E_c}.
\label{eq:char_length}
\end{align}
The resultant length scale is independent of the Kondo-coupling and much shorter than the coherence length $\xi = v_{\rm F}/|V|$. Indeed, we obtain ${\tilde \xi} \simeq v_{{\rm F}\lambda} /E_c = \sqrt{t/|\mu|} \sim 1$ with the use of $E_c  = 2|\mu|$ and $v_{{\rm F}\lambda} = 2\sqrt{|\mu|t}$.

We can understand the appearance of the short length scale independent of the order parameter as follows. 
The length scale of the bound state within the vortex core is characterized by the anomalous self-energy similar to the BCS theory, but here the frequency dependence enters.
In the CMCB-KL, the anomalous self-energy makes the characteristic length be proportional to the frequency as ${\tilde \xi} \sim v_{{\rm F}}/\Delta_\al (\omega) \propto |\omega| v_{{\rm F}}/|V|^2$. 
As a result, the characteristic length becomes very short in the low-energy region $\omega \to 0$ inside the superconducting bulk gap, and the minimal energy is given by the binding energy $\varepsilon_{\al 0} \propto |V|^2/E_c$, to reach Eq.~\eqref{eq:char_length}, where the order parameter dependence is canceled out. 

Considering the boundary condition for $f_{\al \lambda}$, we can determine the coefficient $C_0$ to obtain the quasi-classical Green's function as follows,
\begin{align}
	&g_{\al \lambda}(u) = \pi | \varepsilon_\al (b) | \frac{\mathrm{e}^{- K_{\al \lambda}(u)}}{i\omega_n - \varepsilon_{\al \lambda}(b)}, \\
	\label{eq:energy_spectrum}
	&\varepsilon_\al (b) = - \sg^z_{\al \al} \frac{2|V_\infty|^2}{E_c} + |V_\infty| \frac{b}{v_{{\rm F}\lambda}/2|V_\infty|}. 
\end{align}
See Appendix \ref{sec:energy_spectrum} for detail of the derivation. 
Since the leading-order contribution in the quasiparticle energy $\varepsilon_\al (b)$ is the zeroth-order term of the impact parameter, the characteristic energy scale of the vortex bound state is same as the magnitude of the bulk-gap.

From above, we have clarified that the characteristics of the vortex bound state, such as the short length scale with the order of the lattice constant and the quasiparticle energy with the order of the bulk-gap, are associated with the characteristic frequency dependence of both the normal and anomalous self-energies. 
The quasiparticle energy for the vortex bound state is determined by the normal self-energy, which arises from the effective hybridization between the conduction electron and the pseudofermion. 
On the other hand, the quasiparticle in the low-energy region experiences effectively large pair potential due the frequency dependence of the anomalous self-energy. As a consequence, the characteristic length becomes the atomic scale.

In this section we have assumed the quasiclassical limit $k_{\rm F} \xi \gg 1$ to study the low-energy properties of the vortex bound state.
Whereas the appearing small length scale is not compatible with this assumption, if we regard it as an extrapolation from the quasiclassical limit, a qualitatively same behavior as the two-dimensional tight-binding model is obatined for the isolated vortex.
Hence we expect that the present Eilenberger theory can give qualitatively correct results for the non-uniform superconductors.
The further explorations for the more complex systems, such as vortex lattice state and comparison with three dimensional cases, are left as future studies.

\section{Summary and Discussion}
\label{sec:summary}
Based on the mean-field theory, we have elucidated the physical properties of the vortex bound state in the Kondo lattice model with compensated metallic conduction bands.
We have solved the BdG equation numerically to obtain the self-consistent solution in the presence of the topological defect and the non-topological defect. 
We have revealed that the characteristic length within the vortex core in the CMCB-KL is not sensitive to the choice of the parameters and becomes atomic scale.
This is contrast to the characteristic length in the presence of the impurity potential, which becomes longer with changing parameters such as the Kondo coupling.
Hence the robust short length scale is a characteristic of the vortex state of the CMCB-KL. 
We have also calculated the peak-position of the local DOS to show that the magnitude of the quasiparticle energy is same order as the bulk-gap unlike the BCS superconductor. 

To clarify the physical origin of the characteristics of the vortex bound state, we have constructed the low-energy effective theory of the superconducting state in the CMCB-KL. 
We introduce the low-energy effective Hamiltonian with continuum approximation, where the compensated metallic conduction bands are described by the parabolic dispersion. We then derive the equation of motion of the corresponding Green's function. 
With the use of the quasiclassical approximation, we have derived the Eilenberger equation for the CMCB-KL, where we have the characteristic frequency dependence of the self-energy, which is proportional to $\omega^{-1}$ indicating the odd-frequency superconductivity. 
The validity of the effective theory is checked by comparing it with the tight-binding model and with the bulk properties which are derived without using the quasiclassical theory. 
We study the vortex core bound state of the CMCB-KL using the Kramer-Pesch approximation, which is a perturbative approach originally introduced to describe the vortex bound state in the BCS superconductor. 
As a result, we have revealed that the diagonal self-energy determines the characteristic energy scale where the vortex bound state appears, 
while the anomalous self-energy (pair-potential) is effectively enhanced in the low-energy region to make the length scale of the vortex core very short. 
Thus the peculiar properties of the vortex core is closely related to the dynamical structure of self-energies.

Finally, let us comment on the merit of the vortex core in the CMCB-KL different from the usual BCS case.
It has been recognized that 
the $s$-wave BCS superconductor of the metallic state with the band inversion around the Fermi energy 
has a pair of localized Majorana state at the edges of the vortex line under the magnetic field \cite{Hosur11,G_Xu16}.
In view of the energy spectrum, the zero-energy Majorana state is formed inside the gap with energy $E_{\rm gap} \sim \Delta^2 / E_{\rm F}$ which is a level spacing between the vortex core bound states.
For usual superconductors, this energy $E_{\rm gap}$ is very small due to the magnitude relation $\Delta \ll E_{\rm F}$, and such Majorana mode is observed in the relatively large-$\frac{\Delta}{E_{\rm F}}$ superconductor such as Fe(Se,Te) \cite{Wang18,Zhang18,Machida19}. 
On the other hand, the energy gap in the CMCB-KL that separates zero energy state from the first excited states is the order of the bulk gap, which is much larger than the BCS case. 
Hence, if we consider the topologically non-trivial normal metal in the CMCB-KL, we can expect a zero-energy Majorana mode which is separated from the excited states with the energy of nearly bulk gap and should be easier to be detected experimentally. 
Exploration of such topological superconductors in three dimensions are interesting future perspective in the context of this paper.

\section*{ACKNOWLEDGMENTS}
This work was supported by the Japan Society for Promotion of Science (JSPS) KAKENHI Grants No. 18K13490, No. 18H01176, No. 18H04305, and No. 19H01842. 

\appendix

\section{BCS theory}
\label{sec:BCS}
We use the attractive Hubbard model in the square lattice, which is given by
\begin{align}
	{\mathcal H}_{\rm BCS} &= -t \sum_{\langle i,j \rangle,\sg} \left( c_{i\sg}^\dagger c_{j\sg} + \mathrm{H.c.} \right) - \mu\sum_{i\sg} n_{i\sg} \nonumber \\
	&\hspace{1em}+ U \sum_{i} n_{i\uparrow} n_{i\downarrow},
\end{align}
where $\langle i,j\rangle$ runs over the nearest neighbor bond in the square lattice. $U<0$ denotes the onsite attractive interaction. $n_{i\sg}=c_{i\sg}^\dagger c_{i\sg}^{}$ is the particle number operator of the conduction electron. 
Introducing the spin-singlet Cooper pair amplitude $\Delta_i= U \langle c_{i\downarrow} c_{i\uparrow} \rangle $, the interaction term can be decoupled as follows,
\begin{align}
	U \sum_i n_{i\uparrow} n_{i\downarrow} \simeq \sum_i \Delta_i c_{i\uparrow}^\dagger c_{i\downarrow}^\dagger + \mathrm{H.c.} + \mathrm{const.}.
\end{align}
We thus self-consistently solve the BdG Hamiltonian, which is obtained as follows,
\begin{align}
	&{\mathcal H}_{BdG} = {\vec \Psi}^\dagger {\hat {\mathcal H}} {\vec \Psi} + \mathrm{const.},\\
	&{\hat {\mathcal H}} = \left(
	\begin{array}{cc}
	{\hat \xi} & {\hat \Delta} \\
	{\hat \Delta}^\dagger & - {\hat \xi}^T
	\end{array}
	\right)
\end{align}
where ${\vec \Psi} = (\begin{array}{cc} {\vec c}_\uparrow^{ } & {\vec c}_\downarrow^{~\dagger} \end{array} )^T$ is the Nambu basis with ${\vec c}_\sg = (c_{1\sg},c_{2\sg},\cdots,c_{N\sg} )^T$. 
The matrix elements are given by $({\hat \xi})_{ij} = t_{ij} - \mu\delta_{ij}$ and $({\hat \Delta})_{ij} = \Delta_i \delta_{ij}$.

\section{Current density operator}
\label{sec:current}
In this section, we derive the current density operator for the conduction electrons with the effective mass of the kinetic energy. 
To this end, we utilize the Peierls substitution for the general tight-binding model. In the presence of the electromagnetic field, the Hamiltonian is given as follows,
\begin{align}
	&{\mathcal H}_{{\bm A}} = \sum_{ij,\al \al^\prime} c_{i\al}^\dagger t_{i\al,j\al^\prime} \mathrm{e}^{i \phi_{ij}[{\bm A}] } c_{j\al^\prime}^{ },\\
	&\phi_{ij} [{\bm A}]= e \int^{{\bm R}_i}_{{\bm R}_j} d {\bm r} \cdot {\bm A}({\bm r}),
\end{align}
where $i,j$ is the site index. $\alpha,\alpha^\prime$ denote the physical degrees of freedom such as spin and orbital. $t_{i\al,j\al^\prime}$ is the hopping amplitude. 
$e<0$ is the charge of the electron. $\phi_{ij}$ is the Peierls phase. 
We assume that the wave-length of the electromagnetic field is much longer than the lattice constant for a simplicity. 
Then the Peierls phase is rewriteen as $\phi_{ij} [{\bm A}] \simeq e {\bm A} \left( {\bm R}_{ij} \right) \cdot \left( {\bm r}_{ij}\right)$, where ${\bm r}_{ij} = {\bm R}_i - {\bm R}_j$ and ${\bm R}_{ij} = ({\bm R}_i+{\bm R}_j)/2$ respectively represent the relative coordinate and the center of mass of the conduction electrons. 
In the linear response of the vector potential, the current density ${\hat j}^\mu ({\bm R}_{ij})= -\delta {\mathcal H}_{{\bm A}} /\delta A^\mu ({\bm R}_{ij})$ is obtained as follows,
\begin{align}
	& {\hat j}^\mu ({\bm R}_{ij}) \nonumber \\
	& = - i e \sum_{\al \al^\prime}  c_{i\al}^\dagger \left({\bm r}_{ij} \right)_\mu t_{i\al,j\al^\prime} c_{j\al^\prime}^{ } \nonumber \\
	&+ e^2 \sum_{\nu,\al , \al^\prime}c_{i\al}^\dagger \left({\bm r}_{ij} \right)_\mu \left({\bm r}_{ij} \right)_\nu t_{i\al,j\al^\prime} A^\nu \left( {\bm R}_{ij} \right) c_{j\al^\prime}^{ }.
\end{align}
When we assume the translational symmetry of the original Hamiltonian, the hopping matrix is given by $t_{i\al,j\al^\prime} = t_{\al,\al^\prime} ({\bm r}_{ij})$. 
Then, we can obtain the Fourier component of the current density as follows,
\begin{align}
& {\hat j}^\mu ({\bm q}) = \frac{1}{N} \sum_{i,j} {\hat j}^\mu ({\bm R}_{ij}) \mathrm{e}^{-i {\bm q}\cdot {\bm R}_{ij}} \equiv {\hat j}^\mu_p ({\bm q}) + {\hat j}^\mu_d ({\bm q}),\\
&{\hat j}^\mu_p ({\bm q}) = \frac{1}{N} \sum_{{\bm k}}\sum_{\al\al^\prime} c_{{\bm k}-\frac{{\bm q}}{2}  \al}^\dagger \left( e \frac{ \partial \xi_{{\bm k},\al\al^\prime}}{\partial k_\mu} \right) c_{{\bm k}+\frac{{\bm q}}{2} \al^\prime}^{ } \nonumber \\
&{\hat j}^\mu_d ({\bm q})  \nonumber \\
&= -\frac{1}{N^2} \sum_{{\bm k}{\bm q}^\prime} \sum_{\nu,\al \al^\prime} c_{{\bm k}-\frac{{\bm q}^\prime}{2} \al}^\dagger \left( e^2 \frac{\partial^2 \xi_{{\bm k},\al \al^\prime}}{\partial k_\mu \partial k_\nu} \right) A^\nu ({\bm q}-{\bm q}^\prime) c_{{\bm k}+\frac{{\bm q}^\prime}{2} \al^\prime},
\end{align}
where ${\hat j}^\mu_p ({\bm q})$ and ${\hat j}^\mu_d ({\bm q})$ respectively represent the paramagnetic component and the diamagnetic component. 
$N$ is the number of the site. $c_{{\bm k}\al}$ is the Fourier component of the annihilation operator $c_{i\al}$ defined as $c_{{\bm k}\al}=\sum_i c_{i\al} \mathrm{e}^{-i{\bm k}\cdot {\bm R}_i} /\sqrt{N}$. $\xi_{{\bm k},\al \al^\prime}$ denotes the kinetic energy of the conduction electrons and is given by
\begin{align}
	\xi_{{\bm k},\al \al^\prime} = \sum_{{\bm \delta}} t_{\al \al^\prime} ({\bm \delta}) \mathrm{e}^{-i {\bm k}\cdot {\bm \delta}}.
\end{align}
When we assume that the kinetic energy is diagonal with respect to the channel $\alpha$ and the Fermi energy is located in the bottom (or top) of the conduction band, the energy dispersion can be approximated as  $\xi_{{\bm k},\al \al} = {\bm k}^2/2 m_\al - \mu_\al$. Then, we can rewrite the current density operator as follows,
\begin{align}
	&{\hat j}^\mu_p ({\bm q}) = \frac{1}{N} {\sum_{{\bm k} \al}}^\prime c_{{\bm k}-\frac{{\bm q}}{2} \al}^\dagger \left( \frac{e k_\mu}{m_\al} \right) c_{{\bm k}+\frac{{\bm q}}{2} \al}^{ },\\
	&{\hat j}^\mu_d ({\bm q}) = - \frac{1}{N^2} {\sum_{{\bm k}{\bm q}^\prime,\al}}^\prime c_{{\bm k}-\frac{{\bm q}^\prime}{2} \al}^\dagger \frac{e^2}{m_\al} A_\nu ({\bm q}-{\bm q}^\prime)  c_{{\bm k}+\frac{{\bm q}^\prime}{2} \al},
\end{align}
where the summation $\sum^\prime$ runs over the momentum in the vicinity of the Fermi surface and neglect the high-energy part of the conduction band structure.
In the continuum limit, we replace the field operator $c_{{\bm k}\al}$ with $\psi_{{\bm k}\al}$ whose real-space representation is given by
\begin{align}
	&\psi_\al ({\bm r}) = \int \frac{d{\bm k}}{(2\pi)^\frac{d}{2}} \psi_{{\bm k}\al}^{ } \mathrm{e}^{i {\bm k}\cdot {\bm r}},
\end{align}
where $d$ is the dimension of the system. Then, we can obtain the current density operator in the continuum limit as follows,
\begin{align}
	&{\hat {\bm j}}_p ({\bm r}) = \sum_\al \frac{e}{2 i m_\al} \left( \psi_\al^\dagger ({\bm r}) {\bm \nabla} \psi_\al^{ } ({\bm r}) - \left( {\bm \nabla} \psi_\al^\dagger ({\bm r}) \right) \psi_\al^{ } ({\bm r}) \right),\\
	&{\hat {\bm j}}_d ({\bm r}) = - \sum_\al \psi_\al^\dagger ({\bm r}) \frac{e^2}{m_\al} {\bm A}({\bm r}) \psi_\al^{} ({\bm r}).
\end{align}

\section{Energy dispersion of vortex bound state}
\label{sec:energy_spectrum}
In this section, we summarize the derivation of the energy spectrum in Eq.\eqref{eq:energy_spectrum}. 
To this end, we consider the boundary condition for the anomalous part of the quasi-classical Green's function, which is obtained as follows, 
\begin{align}
	\label{eq:boundary_condition_f}
	&f^{R,A}_{\al \lambda}(\omega,|u|= \infty) \nonumber \\
	&= (-1)^{R,A} \pi \sg^z_{\al \al} \frac{\frac{|V_\infty |^2}{\omega \pm i\delta} \epsilon_{\al {\bar \al}} \mathrm{sgn}(u)}{\sqrt{ -\Lambda_\al^2(\omega \pm i\delta) + \frac{|V_\infty |^4}{(\omega \pm i \delta)^2} }} \nonumber \\
	&\xrightarrow{\omega \to \varepsilon_{\al 0}} (-1)^{R,A} \mathrm{sgn}(u) \pi .
\end{align}
We then obtain the coefficient $C_0$ in the Eq.\eqref{eq:quasiclassical_f_KP} as follows,
\begin{align}
	&C_0 = \frac{\pi v_{{\rm F}\lambda}}{2 W(v_{{\rm F}\lambda})} \frac{1}{\displaystyle z - \frac{1}{2} E_c \sg^z_{\al \al} - \langle\Sigma(z) \rangle + \langle {\bar \Delta_\al}^\prime (z) \rangle }, \\
	&W(v_{{\rm F}\lambda}) = \int_0^\infty du~\mathrm{e}^{-K_{\al \lambda}(u)}, \\
	&\langle \Sigma(z) \rangle = \frac{1}{W(v_{{\rm F}\lambda})} \int_0^\infty du~ \frac{|V(u)|^2}{z} \mathrm{e}^{-K_{\al \lambda}(u)}, \\
	&\langle {\bar \Delta_\al}^\prime (z) \rangle = \frac{1}{W(v_{{\rm F}\lambda})} \int_0^\infty du \frac{b}{|u|} \frac{|V(u)|^2}{z}\epsilon_{\al {\bar \al}}  \mathrm{e}^{-K_{\al \lambda}(u)}.
\end{align}
Since the spatial profile of the order parameter $|V(u)|$ can be regarded as constant as shown in Fig.~\ref{fig:mean_fields_vortex}(a), we put $|V(u)|\simeq |V_\infty|$ to obtain 
\begin{align}
	&K_{\al \lambda}(u)\simeq \frac{|u|}{v_{{\rm F}\lambda}/E_c}, \\
	&W(v_{{\rm F}\lambda}) \simeq \frac{v_{{\rm F}\lambda}}{E_c},\\
	&\langle \Sigma (z) \rangle \simeq \frac{| V_\infty |^2}{z},
\end{align}
In addition, $\langle {\bar \Delta_\al}^\prime (z) \rangle$ is obtained as follows,
\begin{align}
	\langle {\bar \Delta_\al}^\prime (z) \rangle & \simeq \frac{|V_\infty|^2}{z} \sg^z_{\al \al} \frac{E_c}{v_{{\rm F}\lambda}} \int_0^\infty du \frac{b}{u} \mathrm{e}^{- u/(v_{{\rm F}\lambda}/E_c)} \nonumber 
\end{align}
Although this integration shows a logarithmic divergence in $u\to0$, it is an artifact resulting from the perturbative expansion for the impact parameter. 
The factor $1/u$ arises from the pair potential in Eq.\eqref{eq:pair_potential_KP}, where we have assumed $u\gg b \gtrsim0$ to use $1/\sqrt{u^2 + b^2}\simeq 1/u$. 
We therefore introduce the lower cut-off $b$ and ignore the contribution from the small $u$ region since the factor $1/\sqrt{u^2 + b^2}\sim 1/b ~(u< b)$ is regarded as constant. 
In addition, we also introduce the upper cut-off $v_{{\rm F}\lambda}/E_c$ to neglect the exponential term in the integration.
Then, $\langle {\bar \Delta_\al}^\prime (z) \rangle$ is evaluated as follows,
\begin{align}
	\langle {\bar \Delta_\al}^\prime (z) \rangle &\simeq \frac{|V_\infty|^2}{z} \sg^z_{\al \al}  \frac{b}{v_{{\rm F}\lambda}/E_c} \mathrm{log} \left( \frac{v_{{\rm F}\lambda}/E_c}{b} \right) \nonumber \\
	&\sim \frac{|V_\infty|^2}{z}  \sg^z_{\al \al}  \frac{b}{v_{{\rm F}\lambda}/E_c}, 
\end{align}
where we have neglected the coefficient. 
We can find that the denominator of the coefficient $C_0$ is same as $\Lambda_\al (z)$ with $|V_\infty|^2 \to |V_\infty|^2 \left( 1 - \sg^z_{\al \al} b/(v_{{\rm F}\lambda}/E_c)\right)$. 
We then obtain
\begin{align}
	C_0 \simeq \frac{\pi v_{{\rm F}\lambda}}{2 W(v_{{\rm F}\lambda})} \frac{z}{(z - \varepsilon_\al (b))(z - \frac{1}{2} E_c \sg^z_{\al \al})},
\end{align}
where 
\begin{align}
	\varepsilon_\al (b) & = - \sg^z_{\al \al} \frac{2|V_\infty|^2}{E_c} + 2 |V_\infty| \frac{b}{v_{\lambda \perp}/|V_\infty|}
\end{align}
represents the energy dispersion of the vortex bound state. 
Focusing on the low-energy region $|z| \sim |\varepsilon_\al (b)| \ll E_c$, we can rewrite $C_0$ as,
\begin{align}
	C_0 \simeq \frac{\pi |\varepsilon_\al (b)|}{z- \varepsilon_\al (b)}.
\end{align}
Therefore, we can obtain the quasi-classical Green's function as follows
\begin{align}
	&g_{\al \lambda}(u,i\omega_n) = \pi |\varepsilon_\al (b)| \frac{\mathrm{e}^{- K_{\al \lambda}(u)}}{i\omega_n - \varepsilon_\al (b)}.
\end{align}

%

\end{document}